\documentclass[12pt]{article}

\usepackage{amscd, amsmath}
\usepackage{amsthm, amssymb}
\newcommand{\C}{\mathbb{C}}
\newcommand{\R}{\mathbb{R}}
\newcommand{\D}{\partial}
\newcommand{\Db}{\bar{\partial}}
\newcommand{\dd}{{\rm d}}
\newcommand{\M}{{\mathcal M}}

\newtheorem{thm}{Proposition}[section]
\newtheorem{cor}[thm]{Corollary}
\newtheorem{lem}[thm]{Lemma}
\newtheorem*{lema1}{Lemma A.1}
\newtheorem*{lema3}{Lemma A.3}
\newtheorem*{propa2}{Proposition A.2}
\newtheorem*{propa4}{Proposition A.4}

\theoremstyle{remark}
\newtheorem*{rem}{Remark}

\theoremstyle{remark}
\newtheorem{exa}[thm]{Example}

\theoremstyle{remark}
\newtheorem*{que1}{{\bf Question 1}}

\theoremstyle{remark}
\newtheorem*{que2}{{\bf Question 2}}

\theoremstyle{remark}
\newtheorem*{que3}{{\bf Question 3}}

\begin{document}

\begin{titlepage}
\title{
\vskip -70pt
\begin{flushright}
{\normalsize \ DAMTP-2003-59}\\
\end{flushright}
\vskip 25pt
{\bf Some special K\"ahler metrics on $SL(2,\C)$ \\ and their
holomorphic quantization}
}
\vspace{1.4cm}
{\makeatletter
\author{{J. M. Baptista}\thanks{e-mail address: jmqfdb2@damtp.cam.ac.uk}\\
{\small {\sl Department of Applied Mathematics and Theoretical Physics}}\\
{\small {\sl University of Cambridge}} \\
{\small {\sl Wilberforce Road, Cambridge CB3 0WA, England}}
}
\makeatother}
\date{June  2003}
\maketitle
\thispagestyle{empty}
\vspace{1cm}
\vskip 20pt
{\centerline{{\large \bf{Abstract}}}}
\vspace{.35cm}
 The group $SU(2)\times SU(2)$ acts naturally on $SL(2,\C)$ by
simultaneous right and left multiplication. We study the K\"ahler metrics
invariant under this action using a global K\"ahler potential. The volume
growth and various curvature quantities are then explicitly computable.
Examples include metrics of positive, negative and zero Ricci curvature,
and the 1-lump metric of the $\C P^1$-model on a sphere.

  We then look at the holomorphic quantization of these metrics, where
some physically satisfactory results on the dimension of the Hilbert space
can be obtained. These give rise to an interesting geometrical
conjecture, regarding the dimension of this space for general Stein
manifolds in the semi-classical limit.

\end{titlepage}

\part{}

\section{Introduction}

Among the geometrical procedures for quantization, holomorphic
quantization is a particularly simple and natural one, and can be used
whenever the classical system ``lives'' on a complex K\"ahler
manifold. In this paper the complex manifold under study will be
$SL(2,\C)$, and we will consider the K\"ahler metrics on this manifold
which are invariant under a natural action of the group $SU(2) \times
SU(2)$, namely the action defined by simultaneous right and left
multiplication of the matrix in $SL(2,\C)$ by the matrices in $SU(2)$.

$\ $

In the first part of the paper a purely classical study of these
K\"ahler metrics is carried out. We find that each of these metrics
has a global invariant K\"ahler potential, which is essentialy unique,
and is in fact a function of only one real variable. We then use this
potential to compute explicitly several properties of the K\"ahler
manifold. These include the scalar curvature, a potential for the Ricci
form, the volume and volume growth, the geodesic distance from the
submanifold $SU(2) \subset SL(2,\C)$, and a simple criterion for
completeness. Choosing particular functions as K\"ahler potentials we
get metrics with positive-definite, negative-definite and zero Ricci
tensor; the Ricci-flat one being just the usual Stenzel metric on
$T^{\ast}S^3 \simeq SL(2, \C)$.

A significant application of the above results, which was in fact the
original motivation for this paper, is a closer study of the $L^2$-metric
on the moduli space of one lump on a sphere. These lumps are a
particular kind of soliton that appear in $\C P^1$-sigma models, and
have been widely studied \cite{S-Z, Wa}. In particular,
the special case of one 
lump on a sphere has been studied by Speight in \cite{Sp1, Sp2},
where the author also examines general invariant K\"ahler metrics on
$SL(2,\C)$ and finds some of the results mentioned above. The approach
there however is 
rather different, since it is based on the choice of a
particular frame for $T^{\ast} SL(2,\C)$, instead of using the perhaps
more natural K\"ahler potentials.

$\ $

The second part of the paper examines some aspects of holomorphic
quantization on the manifold $SL(2, \C)$ with the K\"ahler metrics
described above. We basically look at two things: the nature and
dimension of the quantum Hilbert space, and the quantum operators
corresponding to the classical symmetries of the metric.

Regarding the latter point, we start by finding the moment map of
the $SU(2)\times SU(2)$ action. This map encodes the classical
symmetries of the system and, through the usual prescriptions of
geometric quantization, subsequently enables us to give an explicit
formula for the operators corresponding to these symmetries. Regarding
the first point, i.e. the dimension of the Hilbert space, the story is
a bit more involved, and we will now spend a few lines describing the
motivation and the results.

If you apply holomorphic quantization to a compact K\"ahler
2$n$-manifold, 
it is a consequence of the Hirzebruch-Riemann-Roch formula that the
dimension of the Hilbert space is finite and
grows asymptotically as $\Omega /(2 \pi \hbar)^n$ when $\hbar
\rightarrow 0$, where $\Omega$ is the volume of the manifold. This
result is also physically interesting, since it agrees with some
predictions of semi-classical statistical mechanics. Trying to see
what happens on the non-compact $SL(2,\C)$ with our invariant metrics,
we were thus led to compute the dimension of the Hilbert space. The
results obtained can be summarized as follows.

The Hilbert space ${\mathcal H}_{HQ}$ in our setting is essentially
the space of square-integrable holomorphic functions on $SL(2,\C)$,
where square-integrable means with respect to some metric-dependent
measure on $SL(2,\C)$. Furthermore all these holomorphic functions
can be seen as restrictions of holomorphic functions on $\C^4
\supset SL(2,\C)$. Defining the subspace ${\mathcal H}_{poly}
\subseteq {\mathcal H}_{HQ}$ of the holomorphic functions which are
restrictions of polynomials in $\C^4$, we then find that ${\rm
dim}{\mathcal H}_{poly} \sim \Omega /(2 \pi \hbar)^3 $ as $\hbar
\rightarrow 0 $ whenever both members are finite. The exact dimension
of ${\mathcal H}_{poly}$, which we also compute, depends on the
particular invariant metric one puts on $SL(2,\C)$; its asymptotic
behaviour however does not. This leads us to conjecture that, as in
the compact K\"ahler case, also for general Stein manifolds
(i.e. complex submanifolds of $\C^N$) this
asymptotic behaviour of ${\rm dim}{\mathcal H}_{poly}$ is
``universal'' -- see the discussion of section 8.

\section{The invariant K\"ahler metrics}

  We start by considering the action of the group $G := SU(2) \times
  SU(2)$ on the complex manifold $M := SL(2, \mathbb{C})$ defined by
\begin{gather}
 \psi : G\times M \longrightarrow M \ \ \ , \ \ (U_1 , U_2 , A)
 \mapsto U_1 A U_2^{-1}\ \ .   \label{e2.1}
\end{gather}
  This is clearly a smooth action which acts on $M$ through
  biholomorphisms. A detailed study of $\psi$ and its orbits is done
  in Appendix A. For example one finds there that all the orbits
  except one have real dimension $5$, the exceptional one being $SU(2)
  \subset M$, which has dimension $3$. For the purposes of this
  section, however, it is enough to quote the following result :

\begin{thm}
  Any smooth $G$-invariant function $\tilde{f} : M \rightarrow \mathbb{R}$
  can be written as a composition $f \circ y$, where $y: M \rightarrow
  [0, +\infty [$ is defined by $y(A) = \cosh^{-1}[\frac{1}{2}
  {\rm tr}(A^{\dagger}A)]$, and $f: \mathbb{R} \rightarrow \mathbb{R}$ is a
  smooth even function.
\end{thm}
  
  We are now interested in studying K\"ahler metrics and forms over
  $M$. To begin with, the well-known diffeomorphism $M \simeq S^3
  \times \mathbb{R}^3$ implies that the de Rham cohomology of $M$ and
  $S^3$ are the same. In particular every closed $2$-form on $M$ is
  exact. On the other hand, regarding $\mathbb{C}^4$ as the set of
  $2\times 2$ complex matrices, we have that $M$ is the hypersurface
  given as the zero set of the polynomial $A \mapsto 1 -
  \text{det}A$. Since the derivative of this polynomial is injective
  on the zero set, $M$ is a complex submanifold of $\mathbb{C}^4$. It
  then follows from standard results in complex analysis of several
  variables (see th. $5.1.5 ,\ 5.2.10$ and $5.2.6$ of \cite{Ho})
  that $M$ is a Stein manifold with Dolbeault groups $H^{p,q} (M) =
  0$ (except for $p=q=0$).

  From all this we get the following lemma:
\begin{lem}
  Any closed (1,1)-form $\omega$ on $M$ can be written $\omega =
  \frac{i}{2} \D \bar{\D} \tilde{f}$, where $\tilde{f}$ is a
  smooth function  on $M$. If $\omega$ is real, then $\tilde{f}$ can
  also be chosen real.
\end{lem}
\begin{proof}
  This is just like the usual proof of the local $\D\Db$-lemma.
  As argued above, the closedness of $\omega$ implies its
  exactness, hence $\omega = \dd \psi = \D\psi^{0,1} + \Db \psi^{1,0}$
  for some $\psi \in H^{1}(M,\C)$. Since $\D\psi^{0,1} = \Db
  \psi^{1,0} = 0$ (because $\omega$ is a (1,1)-form) and 
  $H^{1,0}(M) = H^{0,1}(M) = 0$, we have that $\psi^{0,1}= \Db f_1$
  and $\psi^{1,0} = \D f_2$ for some smooth functions $f_j$ on
  $M$. Defining $\tilde{f} = 2i(f_2 - f_1)$ we thus get $\omega =
  \frac{i}{2} \D \Db \tilde{f}$. If $\omega =  \frac{i}{2} \D \bar{\D}
  \tilde{f}$ is real, then $\frac{1}{2}(\tilde{f} + {\rm c.c.})$ is a real
  potential for $\omega$.
\end{proof}
  Having done this preparatory work, we now head on to the main result
  of this section.
\begin{thm}
  Suppose $\omega \in \Omega^{1,1} (M; \mathbb{R})$ is a closed
  $G$-invariant form. Then one can always write $\omega =  \frac{i}{2}
  \D \bar{\D} (f \circ y)$, where $f$ and $y$ are as in
  proposition $2.1$ and $f \circ y$ is smooth. The function $f$ is unique
  up to a constant. Furthermore, the hermitian metric on $M$
  associated with $\omega$ is positive-definite iff $f' > 0$ on $]0,
  +\infty[$ and $f'' > 0$ on $[0, +\infty[$.
\end{thm}

\begin{proof}
  By the previous lemma $\omega =  \frac{i}{2} \D \bar{\D} \tilde{f}$ for
  some $\tilde{f} \in C^{\infty} (M;\mathbb{R})$. Now, for any $g \in G$,
  the $G$-invariance of $\omega$ and the holomorphy of $\psi_g$
  imply that
\[
\omega\, =\, \psi_g ^{\ast}\, \omega \, = \, \psi_g ^{\ast}\, \frac{i}{2}
  \D \bar{\D} \tilde{f}\,  =\, \frac{i}{2}\, \D \bar{\D} (\tilde{f} \circ
  \psi_g) \ .
\]
  Hence by averaging over $g \in G$ if necessary (recall that $G$ is
  compact), one may assume that the potential $\tilde{f}$ is
  $G$-invariant. The first part of the result then follows from
  proposition $2.1$.

  To establish the second part, recall that the associated hermitian
  metric is defined by 
\begin{gather} 
H(\cdot , \cdot)\, =\, \omega ( \cdot , J \cdot) - i\, \omega (\cdot,
\cdot) \ ,
\label{e2.2}
\end{gather}
 where $J$ is the complex structure on $M$. Since both $\omega$ and
 $J$ are $G$-invariant (the last one because $\psi_g$ is
 holomorphic), we conclude that also $H$ is $G$-invariant. 
 Now consider the complex submanifold $\Lambda \subset M$ consisting of the
 diagonal matrices in $M$. It follows from lemma A.1 of 
 Appendix A that $\Lambda$ intersects every orbit of $\psi$. Hence,
 by the $G$-invariance, $H$ is positive-definite on $M$ iff it is
 positive-definite at every point of $\Lambda$. To obtain the
 condition for positiveness over $\Lambda$ we now use a direct
 computation.

  Take the neighbourhood ${\mathcal U} := \{ A \in M : A_{11} \ne 0\}$
  and the complex chart $\varepsilon$ of $M$ defined by
\begin{gather}
\varepsilon : {\mathcal U} \rightarrow \C ^{\ast} \times \C ^2  \quad
, \quad \varepsilon^{-1}(z_1 , z_2 ,z_3 ) = 
\begin{bmatrix}
z_1  &   z_2 \\
z_3  &  \frac{1+ z_2 z_3}{z_1}
\end{bmatrix}  
\label{e2.3}
\end{gather}
Note that $\Lambda \subset {\mathcal U}$ and that $\varepsilon$ is a chart
of $M$ adapted to $\Lambda$. Defining $x(A) = {\rm
tr}(A^{\dagger} A)/2$ we have that $y = \cosh ^{-1}(x)$ and
\[
x \circ \varepsilon^{-1} (z)\, =\, \frac{1}{2} \left( |z_1|^2 +|z_2|^2 +|z_3|^2
+ |1+z_2 z_3 |^2 /|z_1|^2 \right) \ .
\]
A direct calculation using the chain rule now shows that, on a point
diag$(z_1 , z_1 ^{-1}) \,\in \Lambda$, we have
\begin{gather}
\omega \,=\,  \frac{i}{2} \D \bar{\D} (f \circ y) \, =\, \frac{i}{2} \left[ 
\frac{f''(y)}{|z_1|^2}\, \dd z_1 \wedge \dd \bar{z_1} +
\frac{f'(y)}{2\sinh (y)}\, (\dd z_2 \wedge \dd \bar{z_2} + \dd z_3
\wedge \dd \bar{z_3}) \right] 
\label{e2.5}
\end{gather}
and hence
\begin{gather}
H\, =\, \frac{f''(y)}{|z_1|^2}\, \dd z_1 \otimes \dd \bar{z_1} +
\frac{f'(y)}{2\sinh (y)}\, (\dd z_2 \otimes \dd \bar{z_2} + \dd z_3
\otimes \dd \bar{z_3}) \ \ .   \label{e2.4}
\end{gather}
Thus at points of $\Lambda$ such that $y>0$ (i.e. $|z_1|\ne 1$), we
have $\sinh (y) >0$ and the positive-definiteness of $H$ is equivalent
to $f'(y),\, f''(y) > 0$. On the other hand, since $H$ and the chart
are defined over all of $\Lambda$, continuity implies that at a
point of $\Lambda$ with $y=0$ (i.e. $|z_1| =1$) we must have
\[
H\, =\, f''(0)\, \dd z_1 \otimes \dd \bar{z_1} + \frac{1}{2} f''(0)\, (\dd z_2
\otimes \dd \bar{z_2} + \dd z_3 \otimes \dd \bar{z_3}) \ ,
\]
where it was used that 
\[
\lim_{y\rightarrow 0^+}
\frac{f'(y)}{\sinh (y)}\, =\, \lim_{y\rightarrow 0^+}
\frac{f'(y)}{y}\, =\, f''(0)  \ . 
\] 
Thus at this point the
positive-definiteness of $H$ is equivalent to $f''(0) >0$. This
establishes the last part of the proposition.

 To end the proof we finally note that formula (\ref{e2.4}) implies the
 uniqueness of $f'(y)$, and hence the uniqueness of $f$ up to a constant.
\end{proof}

  Roughly speaking, this proposition guarantees the existence of
  $G$-invariant potentials for $G$-invariant K\"ahler forms. A
  particular feature of these potentials, which will be crucial for
  the explicit calculations later on, is that they are entirely
  determined by their values on the diagonal matrices, since every
  orbit of the $G$-action contains one of these. Having this in mind,
  we now end this section by presenting a technical lemma which will
  prove useful on several occasions.
\begin{lem}
  Suppose $\tilde{f}$ is a smooth $G$-invariant function on $M$, and
  consider the submanifold $\Lambda = \{ {\rm diag}(z_1 , z_1 ^{-1})\, :\,
  z_1 \in \C ^{\ast} \}$ of diagonal matrices in $M$. If $h =
  h(|z_1|)$ is a smooth function on $\Lambda$ such that  $\D \Db h =
  \D \Db \tilde{f} |_{\Lambda}$, then $2\tilde{f}(z_1) = h(z_1) +
  h(z_1 ^{-1}) + {\rm const.}$ on the submanifold $\Lambda$.
\end{lem}
\begin{proof}
  The hypothesis is that $\frac{\D ^2 h}{\D z_1 \D \bar{z}_1 } =
  \frac{\D ^2 \tilde{f}}{\D z_1 \D \bar{z}_1}$ on $\Lambda$. Writing $z_1 \in
  \C ^{\ast}$ as $z_1 = r e^{i\theta}$ and using the expression for
  the laplacian in polar coordinates, we have
\[
0\, =\, \frac{\D ^2 (\tilde{f} -h)}{\D z_1 \D \bar{z_1} }\, =\, \frac{1}{4}
\Delta (\tilde{f} - h)\, =\, \frac{1}{4} (\frac{\D ^2}{\D r^2} +
\frac{1}{r} \frac{\D}{\D r} + \frac{1}{r^2} \frac{\D ^2}{\D \theta
^2}) (\tilde{f} - h)\ .
\]
But the $G$-invariance implies that $\tilde{f}$ only depends on $r$;
since the same is assumed for $h$, we get 
\[
(\frac{\D ^2}{\D r^2} + \frac{1}{r} \frac{\D}{\D r}) (\tilde{f} -h ) =
0\ \ \Rightarrow \ \ \tilde{f}-h = A\log r + B \ .
\]
Now, $G$-invariance also implies that $\tilde{f}(z_1) = \tilde{f}(z_1
^{-1})$, thus
\begin{align*}
2\tilde{f}(z_1)\, & =\, \tilde{f}(z_1) + \tilde{f}(z_1 ^{-1})\, =\, h(z_1) +
h(z_1 ^{-1}) + A(\log |z_1| + \log |z_1|^{-1}) + 2B\, = \\
& =\, h(z_1) + h(z_1^{-1}) + 2B \ .  
\end{align*} 
\end{proof}

\section{Curvature and completeness}

 Throughout this section $\omega$ will be the K\"ahler form of a $G$-invariant
K\"ahler metric on $M$. Thus according to proposition $2.3$ we can write
\begin{gather}
\omega \, =\,  \frac{i}{2}\, \D \bar{\D} (f \circ y) \ ,     
\label{e3.1}
\end{gather}
where $f \circ y$ is smooth and $f$ satisfies all the conditions of
proposition $2.3$.

  The first task now is to calculate the Ricci form $\rho$ associated
  to this K\"ahler metric. More precisely, we will obtain a potential
  for $\rho$ expressed in terms of the function $f$.
\begin{thm}
The Ricci form of the metric with K\"ahler form $\omega$ is given by
\[
 \rho\ =\ -i\,\D \Db \log \left[ \left(\frac{f'(y)}{\sinh(y)}\right)^2 \,f''(y)
 \right]\ .
\]
\end{thm}
\begin{proof}
 The $G$-invariance of the metric implies the $G$-invariance of the
 Ricci form $\rho$. Thus, by proposition $2.3$, $\rho$ has a global
 $G$-invariant potential $\tilde{\rho}$. Now consider the chart
 $({\mathcal U}, z_1 ,z_2 ,z_3 )$ for $M$ defined in the proof of the
 same proposition. According to a standard result, if in this chart
\[
\omega |_{\mathcal U}\, =\, \frac{i}{2}\, h_{\alpha \, \bar{\beta}}\, \dd
 z^{\alpha} \wedge \dd \bar{z}^\beta \ , 
\] 
then the Ricci form is given by
\[
\rho |_{\mathcal U}\, =\, -i \D \Db \log (\det h_{\alpha \, \bar{\beta}})\ .
\]
In particular, over the complex submanifold $\Lambda$ of diagonal
matrices we have 
\[
\frac{i}{2}\, \D \Db \tilde{\rho}\,|_{\Lambda}\, =\,
\rho\,|_{\Lambda}\, =\,  -i \D \Db \log (\det h_{\alpha
\bar{\beta}})\, |_{\Lambda}\ .
\]
But (\ref{e2.4}) gives us $ h_{\alpha \bar{\beta}}$ over $\Lambda$, and so we
compute that
\[
\log (\det h_{\alpha \bar{\beta}})\, |_{\Lambda}\, =\, \log
\left(\frac{1}{|z_1|^2} \left(  \frac{f'(y)}{2\sinh y} \right)^2
f''(y)  \right)\ .
\]
Since this function only depends on $|z_1|$, by lemma $2.4$ we get that 
\[
\tilde{\rho} |_{\Lambda}\, =\, -2\log \left( \left(\frac{f'(y)}{\sinh y}
\right)^2 f''(y) \right) + {\rm const.} \ \ .
\]
Finally the $G$-invariance of $\tilde{\rho}$ guarantees that this
expression is valid all over $M$. Thus we conclude that $\rho =
\frac{i}{2} \D \Db \tilde{\rho}$ has the stated form.
\end{proof}

The next step is the computation of the scalar curvature. Note that
the $G$-invariance of the metric implies the $G$-invariance of this
function.
  
\begin{thm}
The scalar curvature of the Riemannian metric associated with the
K\"ahler form $\omega$ is 
\[
s\, =\, \frac{2}{f'' (f')^2} \frac{\dd}{\dd y} \left( (f')^2
\frac{\dd}{\dd y} \log \left( \frac{\sinh ^2 y}{f'' (f')^2} \right)
\right)\ .
\]
\end{thm}
\begin{proof}
Let us call $g(y) := \log \left( \frac{\sinh ^2 y}{f'' (f')^2}
\right)$, so that $\rho = i \D \Db (g \circ y)$. The same calculations
that led to formula (\ref{e2.5}) now give
\begin{gather}
\rho |_{\Lambda}\, =\, i \left( \frac{g''}{|z_1|^2}\, \dd z_1 \wedge \dd
\bar{z_1} + \frac{g'}{2\sinh y} (\dd z_2 \wedge \dd \bar{z_2} + \dd
z_3 \wedge \dd \bar{z_3}) \right) \ .  \label{e3.2}
\end{gather}
Writing $\omega  = \frac{i}{2} h_{\alpha \, \bar{\beta}}\, \dd
 z^{\alpha} \wedge \dd \bar{z}^\beta$ and $\rho  = \frac{i}{2}\,
 r_{\alpha \, \bar{\beta}}\, \dd z^{\alpha} \wedge \dd \bar{z}^\beta$,
the scalar curvature of the associated Riemannian metric is $s = 2
 h^{\alpha \, \bar{\beta}} r_{\alpha \, \bar{\beta}}$. Thus using (\ref{e2.5})
and (\ref{e3.2}) we can compute the restriction of $s$ to the submanifold
 $\Lambda$: 
\begin{gather}
s |_{\Lambda}\ =\ 2 \frac{g''}{f''} + 4 \frac{g'}{f'}\ =\ 
\frac{2}{f''(f')^2} \frac{\dd}{\dd y}((f')^2 g') \ .
\end{gather}
The $G$-invariance of $s$ then shows that this formula is valid all
over $M$.
\end{proof}

$\ $

In the last part of this section we will make contact with a paper by
Patrizio and Wong \cite{P-W}: this will give us almost for free some results
about the completeness of the $G$-invariant metric associated to
$\omega$.

To make contact one just needs to note that the linear transformation
on $\C ^4$ defined by the matrix 
\[
\begin{bmatrix}
1  & 0 & 0 & -i \\
0  & 1 & -i& 0  \\
0  & -1& -i& 0  \\
1  & 0 & 0 & i
\end{bmatrix}
\]   
takes the standard hyperquadric $Q_4 = \{ w\in \C ^4 : \sum w_k ^2 \,
=1\}$ to $M$, and the norm function $\| w \|^2$ on $Q_4$ to the
function $x(A)= {\rm tr}(A^{\dag}A)/2$ on $M$. Therefore all
the results in \cite{P-W} valid for $(Q_n , \| w\|^2)$ can be restated here
for $(M,x)$. In particular we have that
\begin{itemize}

\item[(1)] The function $y = \cosh ^{-1} x$ is plurisubharmonic
exhaustion on $M$, and solves the homogeneous Monge-Amp\`ere equation
on $M - y^{-1}(0) = M - SU(2)$ (\cite{P-W}, th. 1.2). 

\item[(2)]Suppose $\tilde{f} = f \circ y$ is a strictly
plurisubharmonic function on $M$. Then with respect to the metric
defined by $\frac{i}{2} \D \Db \tilde{f}$, the distance in $M$ between
the level sets $\{ y=a\}$ and $\{ y=b \ge a\}$ is (\cite{P-W}, th. 3.3)
\begin{equation}
D(a,b)\, =\, \frac{1}{\sqrt{2}} \int_{f(a)}^{f(b)} \sqrt{-
\frac{(f^{-1})''(t)}{(f^{-1})'(t)}}\, \dd t\, =\, \frac{1}{\sqrt{2}}
\int_{a}^{b} \sqrt{f''(y)} \,\dd y \ .         
\label{e3.3}
\end{equation}
Furthermore, the distance-minimizing geodesics between these level
sets are the integral curves of the vector field $X / \|X\|$, where
$X$ is the gradient vector field of $\tilde{f}$ (one can check
directly that $X \ne 0$ on $M-SU(2)$). 

\end{itemize}
As a consistency check, we remark that the strict plurisubharmonicity
of $\tilde{f} = f \circ y$ together with proposition $2.3$ garantees that
$f''(y) > 0$ on $[0,+\infty [ = y(M)$; thus the integral formula for
the distance is well defined. It is now more or less straightforward
to prove the following proposition. 

\begin{thm}
The metric on $M$ with K\"ahler form $\omega$ is complete if and only
if 
\[
D(0, +\infty)\, =\, \frac{1}{\sqrt{2}} \int_{0}^{+\infty} \sqrt{f''(y)}
\,\dd y \, =\, +\infty \ .
\]
\end{thm}

\begin{proof}
By Hopf-Rinow, the metric is complete iff the closed bounded sets of
$(M,\omega)$ are compact. So suppose
that $D(0, +\infty) = +\infty$ and that $B$ is a closed and
bounded subset of $M$. Then for $b$ big enough we have
\[
D(0,b)\, >\, \sup_{x\in B}\, D(0, y(x))\quad \Rightarrow \quad B\, \subset
\, y^{-1}([0,b]) = x^{-1} ([1, \cosh b])\ .
\]
But $x$ is just the usual norm on $\C^4$ restricted to $M$, thus $B$
is also closed and bounded in $\C^4$, and so is compact.

Conversely, if $D(0, +\infty) < +\infty$, then $M$ itself is a closed
bounded set which is not compact, and thus the metric is incomplete.
\end{proof}

\section{Volume and integration}

The purpose of this section is to study the integrals over
$(M,\omega)$ of $G$-invariant functions, where $\omega$ is as in
(\ref{e3.1}). More precisely, we want to prove the following result.

\begin{thm}
Let $\tilde{h}$ be a smooth $G$-invariant function on $M$, which by
proposition $2.1$ can be written $\tilde{h}= h \circ y$, and let $M_r$ be
the open submanifold $y^{-1}([0,r[) \subset M$. Then we have that
\begin{gather}
\int_{M_r} \tilde{h}\ \frac{\omega ^3}{3!}\  =\  \frac{\pi ^3}{3}
\int_{0}^{r} h(y)\, \frac{\dd}{\dd y} (f'(y))^3 \ \dd y \ .
\end{gather}
\end{thm}

Notice that $\omega ^3 / 3!$ is the volume form of the metric on $M$
associated with $\omega$, so with the particular choice $\tilde{h}
\equiv 1$ we get the volume of $M_r$. Remark also that with $\tilde{h}
\equiv 1$ or $\tilde{h} \equiv s$, where $s$ is the scalar curvature
given by proposition $3.2$, the integral on the right-hand side is
trivially computable. Thus taking into account the restrictions on $f$
imposed by propositions $2.1$ and $2.3$, one gets the following corollary.

\begin{cor}
For the K\"ahler metric on $M$ associated with $\omega$, the volume of
$M_r$ and the integral of the scalar curvature over $M_r$ are,
respectively, 
\[
\frac{1}{3} (\pi f'(r))^3 \qquad {\rm and} \qquad 2\pi^3 \left( f'(y)^2
\frac{\dd}{\dd y} \log \left( \frac{\sinh ^2 y}{f''(y) f'(y)^2}
\right) \right)_{y=r} \ .
\] 
In particular $M$ has finite volume iff $f'(r)$ is bounded.
\end{cor}

We now embark on the proof of proposition $4.1$. To start with, it will
be convenient to restate here some results used in \cite{Sp1, Sp2} to
study the lump metric.

Consider the Pauli matrices
\[
\tau_1 = \begin{bmatrix}
        0  &  1 \\
        1  &  0    \end{bmatrix} \ \ ,\ \ 
\tau_2 = \begin{bmatrix}
        0  &  -i \\
        i  &  0    \end{bmatrix} \ \ ,\ \
\tau_3 = \begin{bmatrix}
        1  &  0 \\
        0  &  -1   \end{bmatrix}  \ \ ,
\]
so that $\{ \frac{i}{2} \tau_a \}$ is a basis for the Lie algebra
$su(2)$. Associated to each $\frac{i}{2} \tau_a$ is a left-invariant
$1$-form $\sigma_a$ on $SU(2)$, and $\{ \sigma_a\}$ is a global
trivialization of the cotangent bundle of $SU(2)$. Then according to
\cite{Sp1, Sp2} and the references therein we have that :
\begin{itemize}

\item[$\bullet$] There is a diffeomorphism $\,\chi : SU(2) \times \R ^3
\rightarrow M\,$ defined by
$$
\chi (U, \vec{\lambda }) = U\, (\sqrt{1+ \lambda ^2 } I +
\vec{\lambda } \cdot \vec{\tau }) \ , \ \ {\rm with}\quad \lambda
= | \vec{ \lambda } | \ .
$$

\item[$\bullet$] The usual action $\psi$ of $G$ on $M$ is taken by $\chi$ to
the action $\tilde{\psi}$ on $SU(2) \times \R^3$ given by
\begin{gather}
\tilde{\psi}_{(U_1 ,U_2)} (U, \vec{\lambda})\ =\ (U_1 U
U_2 ^{-1} ,\ {\mathcal R}_{U_2} (\vec{\lambda}))
\label{e4.1}
\end{gather}
where ${\mathcal R}:SU(2)\rightarrow SO(3)$ is the usual double
covering; explicitly ${\mathcal R}_{U_2} \in SO(3)$ has components 
$({\mathcal R}_{U_2})_{ab} = \frac{1}{2} {\rm tr}(\tau_a U_2 \tau_b U_2
^{\dag})$.

\item[$\bullet$] Regarding the $\sigma_a$ and the $\dd \lambda_a$ as $1$-forms
defined over $SU(2)\times \R^3$, the action $\tilde{\psi}$ acts on
these forms by $\tilde{\psi}^{\ast}_{(U_1 , U_2)} (\vec{\sigma} ,\, \dd
\vec{\lambda}) = ({\mathcal R}_{U_2} \vec{\sigma} ,\, {\mathcal R}_{U_2}
\dd \vec{\lambda} )$.

\item[$\bullet$] The Euler angles $(\beta ,\alpha ,\gamma ) \in\, ]0 ,
4\pi[ \times ]0 , \pi[ \times ]0 , 2\pi[ $ define an oriented chart of
$SU(2)$ with dense domain such that, on this domain,
\begin{align}
\sigma_1 & = -\sin \gamma \ \dd \alpha + \cos \gamma\, \sin \alpha\  \dd
\beta  \nonumber \\
\sigma_2 & = \cos \gamma\  \dd \alpha + \sin \gamma\, \sin \alpha\  \dd
\beta   \label{e4.2}  \\
\sigma_3 & = \cos \alpha\  \dd \beta + \dd \gamma \ .  \nonumber
\end{align} 

\end{itemize}

The plan now is to use the diffeomorphism $\chi$ to compute the
integrals on $SU(2) \times \R^3$, instead of $M$. Since the $\{
\sigma_a , \lambda_a\}$ trivialize the cotangent bundle of  $SU(2)
\times \R^3$, the pull-back by $\chi$ of the volume form on $M$ can be
written 
\[
\mu\ :=\ \chi^{\ast}\, \frac{\omega^3}{3!}\ =\ \hat{\mu} (U , \vec{\lambda}) \ \sigma_1 \wedge \sigma_2 \wedge
\sigma_3 \wedge \dd \lambda_1 \wedge \dd \lambda_2 \wedge \dd
\lambda_3 \ ,
\] 
for some non-vanishing function $\hat{\mu}$ on $SU(2) \times
\R^3$. Moreover, $\mu$ must be invariant under $\tilde{\psi}$, because the
volume form on $M$ is invariant under $\psi$. But notice now
that, under $\tilde{\psi}$,
$$
\vec{\sigma} \mapsto {\mathcal R}_{U_2}\vec{\sigma}  \quad \Rightarrow
\quad \sigma_1 \wedge 
\sigma_2 \wedge \sigma_3\ \mapsto\ \det ({\mathcal R}_{U_2})\, \sigma_1 \wedge
\sigma_2 \wedge \sigma_3 \, =\, \sigma_1 \wedge \sigma_2 \wedge
\sigma_3 \ ,
$$
because {${\mathcal R}_{U_2} \in SO(3)$. For the same reason, also $ \dd
\lambda_1 \wedge \dd \lambda_2 \wedge \dd \lambda_3$ is invariant, and
hence  $\sigma_1 \wedge \sigma_2 \wedge \sigma_3 \wedge \dd \lambda_1
\wedge \dd \lambda_2 \wedge \dd \lambda_3$ is invariant too. This fact
together with the invariance of $\mu$ implies the invariance of the
function $\hat{\mu}$. From the formula (\ref{e4.1}) for the action
$\tilde{\psi}$ it is then clear that $\hat{\mu}$ only depends on $\lambda
= |\vec{\lambda}|$.

The computation of the function $\hat{\mu} (\lambda)$ is now
straightforward. First we have
\begin{align*}
\hat{\mu} (\lambda)\ & =\ \mu_{({\rm Id}, 0,0,\lambda)} \left(\frac{i}{2} \tau_1
,\,\frac{i}{2} \tau_2 ,\, \frac{i}{2} \tau_3 ,\, \frac{\D}{\D \lambda_1},\,
\frac{\D}{\D \lambda_2},\,  \frac{\D}{\D \lambda_3}\right) = \\
& =\ \frac{1}{6}\, (\omega^3)_{\chi ({\rm Id}, 0,0,\lambda)} \left(\chi_{\ast}
(\frac{i}{2} \tau_1) ,\, \ldots ,\, \chi_{\ast} \frac{\D}{\D \lambda_3}\right)\ .
\end{align*}
On the other hand, using the chart (\ref{e2.3}) and (\ref{e2.5}), at the
point $q(\lambda) := \chi ({\rm Id}, 0,0,\lambda) = {\rm
diag}(\sqrt{1+\lambda^2} + \lambda ,\sqrt{1+\lambda^2} - \lambda )$ of
$M$ we also have 
$$
\frac{1}{6}\, (\omega^3)_{q(\lambda)}\ =\ (\frac{i}{2})^3
\frac{f''(y)}{(\sqrt{1+\lambda^2} + \lambda)^2} \left( \frac{f'(y)}{2
\sinh y} \right) ^2 \, \dd z_1 \wedge \dd \bar{z_1} \wedge \dd z_2
\wedge \dd \bar{z_2} \wedge \dd z_3 \wedge \dd \bar{z_3} \ .  
$$
Finally a tedious calculation that we will not reproduce shows that
\[
(\dd z_1 \wedge \dd \bar{z_1} \wedge \dd z_2 \wedge \dd \bar{z_2}
\wedge \dd z_3 \wedge \dd \bar{z_3})_{q(\lambda)} \left(\chi_{\ast}
(\frac{i}{2} \tau_1) , \ldots , \chi_{\ast} \frac{\D}{\D \lambda_3}\right) =
4i \sqrt{1+\lambda^2} (\sqrt{1+\lambda^2} + \lambda)^2 \, , 
\]
and so we get 
$$
\hat{\mu}(U,\vec{\lambda})\, =\, \hat{\mu} (\lambda)\, =\,
\frac{\sqrt{1+\lambda^2}}{8} \left( \frac{f'(y \circ \chi)}{\sinh (y
\circ \chi)} \right)^2 f''(y\circ \chi) \ .
$$
Having calculated the volume form on $SU(2) \times \R^3$, the rest of
the proof of proposition $4.1$ goes on smoothly.

Call as usual $x(A) = {\rm tr}(A^\dag A)/2$ and $y =
\cosh^{-1} (x)$. A quick calculation shows that $x\circ \chi (U,
\vec{\lambda}) = 1+ 2\lambda^2$, and so we have an explicit
relation $y =y(\lambda)$. From this relation it is clear that
$\chi^{-1}(M_r) = SU(2) \times B_l$, where $B_l$ is the open ball,
centered at the origin of $\R^3$, with radius $l$ such that $1+2l^2 =
\cosh r$. Hence, for any invariant function $\tilde{h} = h \circ y$
on $M$ we have 
\begin{align*}
\int_{M_r} \tilde{h}\, \frac{\omega^3}{3!}\ & =\  \int_{\chi^{-1} (M_r)}
(\tilde{h} \circ \chi)\, \mu\ =\ \int_{SU(2) \times B_l} (h\cdot \hat{\mu})(y
(\lambda))\   \sigma_1 \wedge \sigma_2 \wedge \sigma_3
\wedge \dd \lambda_1 \wedge \dd \lambda_2 \wedge \dd \lambda_3 \, \\
&  =\, 
\left( \int_{SU(2)} \sigma_1 \wedge \sigma_2 \wedge \sigma_3 \right)
\int_{0}^{l}  (h\cdot \hat{\mu})(y(\lambda))\, 4\pi \lambda^2 \,\dd
\lambda \ .
\end{align*}
Using the value of $\hat{\mu}(\lambda)$ and the relation $y=y(\lambda)$,
a change of variables in the last integral shows that it coincides
with
$$
\frac{\pi}{16} \int_{0}^{r} h(y)\, f''(y)\, (f'(y))^2 \ \dd y \ .
$$ 
The first integral can be computed using (\ref{e4.2}). Namely we have
\[
 \int_{SU(2)} \sigma_1 \wedge \sigma_2 \wedge \sigma_3 \ =\ 
 \int_{0}^{2\pi} \int_{0}^{\pi} \int_{0}^{4\pi} \sin \alpha \ \dd
 \beta \,\dd \alpha \, \dd \gamma \ =\ 16\pi^2 \ .
\]
Putting these two results together we finally obtain the formula
stated in proposition $4.1$.

\section{Examples}

\subsection{The one-lump metric}

The so-called moduli space of degree $1$ lumps on a sphere, which we
will call ${\mathcal M}$, is just the group of rational maps $S^2
\rightarrow S^2$. Identifying $S^2 \simeq \C P^1$, this group is the
same as the group of projective transformations
$$
PGL(2,\C)\ =\ GL(2,\C)/ \C^{\ast}\ =\ SL(2, \C) / \{\pm 1\} \ .
$$
In the physics literature, ${\mathcal M}$ is the space of
minimal energy static solutions of the sigma-model defined on the
Lorentzian spacetime $S^2 \times \R$ with $S^2$ as target space. The
kinetic energy functional of 
this sigma-model induces a certain Riemannian metric on ${\mathcal
M}$, which is also very natural geometrically. It can be defined in the
following way.

Let $w_t : \C P^1 \rightarrow \C P^1$ be a one parameter family of
projective transformations, i.e. a curve on $\M$, and call $w'_0$ its
tangent vector at $t=0$. For each $x \in \C P^1$, $t\mapsto w_t (x)$
is a curve in $\C P^1$, and we call $v(x) \in T_{w_0 (x)} \C P^1$ its
tangent vector at $t=0$. Then the Riemannian metric $g$ on $\M$ is
defined by 
\begin{gather}
g(w'_0 , w'_0) := \int_{x \in \C P^1} h(v(x), v(x))\ {\rm vol}_h
\end{gather}
where $h$ is the Fubini-Study metric on $\C P^1$ and ${\rm vol}_h$ is
the associated volume form. 
In informal terms, one may say that the squared-length of an
infinitesimal curve $t\mapsto w_t$ in $(\M ,g)$ is just
the average over $x\in \C P^1$ of the squared-lengths of the
infinitesimal curves $t\mapsto w_t (x)$ in
$({\mathbb C}P^1 , h)$; thus the measure of ``displacement'' in $\M$
is how much the image points of $w_t$ are moved. 
Using the fact that transformations in
$PSU(2) \subset PGL(2, \C)$ are isometries of $(\C P^1 ,h)$, it is not
difficult to check that right and left multiplication in $PGL(2,\C)$
by elements of $PSU(2)$ are in fact isometries of $(\M , g)$.

$\ $

Now consider the usual chart of the projective space $\C P^1 \setminus
\{ [0,1]\} \rightarrow \C\ ,\ [1,z]\mapsto z$, and let $(u^1,u^2,u^3)$
be any complex chart of $\M$ defined on a neighbourhood of the point
$w_0$. In these charts we have
\begin{align*}
w'_0 \,&  =\, \frac{\dd u^j}{\dd t}(0)\, \frac{\D}{\D u^j}  \\
v(z)\, & =\, \frac{\dd}{\dd t} w_t (z)\, =\, \frac{\dd}{\dd t} w_{u(t)} (z)\, =
\frac{\D}{\D u^j} \left(w_{u}(z)\right)\, \frac{\dd u^j}{\dd t} (0)\,
\frac{\D}{\D z}  \\
h(\frac{\D}{\D z} , \frac{\D}{\D z})\,& =\, h_{1\bar{1}}\, =\, \frac{\D ^2}{\D
z \D \bar{z}} \log (1 + |z|^2)  \\
{\rm vol}_h \,& =\, \frac{i}{2} \frac{\dd z \wedge \dd \bar{z}}{(1+ |z|^2)^2}
\end{align*}
where the last two equalities are standard properties of the
Fubini-Study metric. Calling $\rho := \log (1+|z|^2)$ the local
potential of the Fubini-Study metric we get
\begin{align*}
g(w'_0 , w'_0)  & = \int_{z\in \C} \frac{\D^2 \rho}{\D z \D \bar{z}} (w_0
(z)) \, \frac{\D (w_u (z))}{\D u^j}\, \frac{\dd u^j}{\dd t}\, \frac{\D
(\bar{w}_u (z))}{\D \bar{u}^k}\, \frac{\dd \bar{u}^k}{\dd t}\,
\frac{i}{2}  \frac{\dd z \wedge \dd \bar{z}}{(1+ |z|^2)^2}\ = \\
 & =\ \frac{\dd u^j}{\dd t}\, \frac{\dd \bar{u}^k}{\dd t}\, \int_{z \in
\C} \frac{\D ^2}{\D u^j \D \bar{u}^k} [\rho (w_u (z))]\ \frac{i}{2}
\frac{\dd z \wedge \dd \bar{z}}{(1+ |z|^2)^2}\ = \\
 & =\ \frac{\dd u^j}{\dd t}\, \frac{\dd \bar{u}^k}{\dd t}\,\frac{\D
^2}{\D u^j \D \bar{u}^k}\ \frac{i}{2} \int_{z\in \C} \rho [w_u (z)]\,
\frac{\dd z \wedge \dd \bar{z}}{(1+ |z|^2)^2} \ . 
\end{align*}
Since this equation is valid in any chart $(u^k)$ of $\M$, we conclude
that the function 
\[
a(w)\ :=\ \frac{i}{2} \int_{z \in \C} \log (1 + |w(z)|^2)\, \frac{\dd z
\wedge \dd \bar{z}}{(1+ |z|^2)^2} 
\]
is a global K\"ahler potential for the K\"ahler form on $\M$
associated with the Riemannian metric $g$. Calling this form $\omega$,
we thus have $\omega = \frac{i}{2} \D \Db a$.

It turns out, however, that the integral defining $a(w)$ is difficult
to compute for a general $w \in PGL(2, \C)$, and so we cannot
calculate the potential directly. To circumvent this obstacle we
proceed in the following way.

Firstly, using the double cover $\pi : SL(2, \C) \rightarrow PGL(2,
\C)$, we work on the more palpable group $SL(2, \C)$. Notice that $\pi
^{\ast} \omega = \frac{i}{2} \D \Db (a \circ \pi)$, because $\pi$ is
holomorphic. Moreover, the invariance of $g$ and $\omega$ by right and
left multiplication by elements of $PSU(2)$, implies that $\pi ^{\ast}
\omega$ is invariant by the usual action $\psi$ of the group $G$ on
$SL(2, \C)$. Thus we are on familiar ground. From proposition $2.3$ we
get that $\pi^{\ast}{\omega}= \frac{i}{2} \D \Db \tilde{f}$, for some
$G$-invariant function $\tilde{f}$. The plan now is to compute
$\tilde{f}$ using the potential $a(w)$ and lemma $2.4$.

In fact, for a diagonal matrix $A = {\rm diag}(\xi , \xi^{-1})$ one
can compute that
\begin{align*}
a\circ \pi (A)\ & =\ \frac{i}{2} \int_{z \in \C} \log (1+ \left|
\frac{z}{\xi^2} \right|^2)\, \frac{\dd z \wedge \dd \bar{z}}{(1+
|z|^2)^2}\ = \\
 & =\ 2\pi \int_{0}^{+\infty} \log (1 + \frac{r^2}{|\xi|^4})\,
\frac{r}{(1+r^2)^2} \ \dd r\ =\ \pi\, \frac{\log |\xi|^4}{|\xi|^4 -1}\ ,
\end{align*}
and since
\[
\D \Db (a\circ \pi)|_{\Lambda}\, =\, -2i(\pi^{\ast}\omega)|_{\Lambda}\, =\, \D
\Db \tilde{f} |_{\Lambda}\ ,
\]
from lemma $2.4$ we get that
\[
2\tilde{f} |_{\Lambda}\, =\, 2\pi\, \frac{|\xi|^4 +1}{|\xi|^4 -1}\, \log
|\xi|^2 \ .
\]
Now using the formulas $x(A) = {\rm tr}(A^{\dag}A)/2  =
(|\xi|^2 +|\xi|^{-2})/2$ and $y = \cosh ^{-1} (x)$, a little
algebra shows that, over $\Lambda \subset SL(2, \C)$,
\begin{gather}
\tilde{f}\ =\ \pi \frac{x}{\sqrt{x^2 -1}} \log (x + \sqrt{x^2 -1})\ =\ \pi
\, y\, \coth y \ .
\label{e5.1}
\end{gather}
The $G$-invariance of $\tilde{f}$ finally guarantees that this formula
is valid all over $SL(2, \C)$. We have thus obtained an explicit
potential for the K\"ahler form $\pi^{\ast}\omega$. Notice that
$\tilde{f}(A) = \tilde{f}(-A)$ for any matrix in $SL(2, \C)$, and so
$\tilde{f}$ descends to a function on $PGL(2,\C )$; this will be a
potential for the K\"ahler form $\omega$ on this space.

Using the potential function $\tilde{f}$ and the results of the
previous sections, we will now derive a series of properties of the
metric $g$. Except for the volume and the Ricci potential
computations, these properties were already obtained in \cite{Sp2},
using different methods.

Substituting expression (\ref{e5.1}) into propositions $3.1$ and $3.2$, we obtain a
potential for the Ricci form and the scalar curvature in $(M ,
\pi^{\ast} \omega)$. The first is 
\[
\tilde{\rho}(y)\, =\, -2\log{(y\cosh y - \sinh y)(\sinh 2y - 2y)^2
/(\sinh y)^9}
\]
and the second has a rather long expression which we will not
transcribe. The plot of this expression, however, coincides
with the one in \cite{Sp2}
\footnote{Actually our scalar curvature is half of the one in
\cite{Sp2}, but this must be due to different conventions.}, 
and thus the scalar curvature is a positive
increasing function of $y$ that diverges at infinity. It is
worthwhile noting that, for this metric, the positiveness of the
scalar curvature actually comes from the positive-definiteness of the
Ricci tensor, as can be seen by applying proposition $2.3$ to the
potential $\tilde{\rho}$. Using the criterion of proposition $3.3$ one
may also easily verify that the metric $g$ is incomplete. 
Finally, from corollary $4.2$, and introducing a factor $1/2$ to account
for the double cover $(M , \pi^{\ast}g) \rightarrow (\M ,g)$, we
obtain that the volume of the moduli space is 
\[
{\rm vol}(\M,\, g)\ =\ \frac{\pi^6}{6} \ .
\]  

\subsection{Other metrics}

We will briefly mention here other examples of $G$-invariant metrics
on $M$; these are interesting for their curvature properties.

First of all it is clear from proposition $3.1$ that any solution of 
\[
\frac{\dd}{\dd y} (f'(y))^3 \ =\ c\,(\sinh y)^2 \quad ,\quad c > 0\ ,
\] 
will give rise to a Ricci-flat metric on $M$. This metric coincides with the
Stenzel metric on $TS^3 \simeq SL(2,\C)$ \cite{St}, as can be seen by
using the correspondence $\M \leftrightarrow Q_4$ described in section
$3$ and comparing with section 7 of \cite{St}. It is a complete
metric.

Experimenting with other even functions $f(y)$ one can find metrics
with a wide range of behaviours. For example it follows from
propositions $3.3$, $3.1$ and $2.3$ that the metrics defined by $f(y)=y^2$ and
$f(y)=\cosh y$ are complete and have, respectively, positive-definite
and negative-definite Ricci tensor. The last one is just the induced
metric by the natural inclusion $M \subset \C^4$. The first one has
also the pleasant property that the parameter $y$ is precisely the
geodesic distance from the submanifold $SU(2) \subset M$, and so the
volume of $M_r$ grows exactly with the cube of this distance (see
(\ref{e3.3}) and corollary $4.2$).

\part{Holomorphic Quantization}

In the second part of the paper we want to study the holomorphic
quantization of the K\"ahler manifolds $(SL(2,\C) , \omega)$, where
$\omega$ is any $G$-invariant K\"ahler form. We will firstly obtain
the quantum operators corresponding to the classical symmetries of the
system. After that we will compute the dimension of the Hilbert space
of the quantized system. This last calculation takes a bit of work,
but in the end we find some physically satisfactory results, as
described in the Introduction.

\section{The classical moment map} 

Recall the action $\psi :G\times M \rightarrow M$ described in section
$1$, and suppose $\omega = \frac{i}{2} \D \Db (f\circ y)$ is any
$G$-invariant K\"ahler form on $M$ (see proposition $2.3$). Then,
tautologically, $\psi$ is a symplectic action on $(M,\omega)$. Since
$G$ is a compact semi-simple Lie group, general results state that
there is a unique moment map $\mu : M\rightarrow {\mathfrak g}^{\ast}$
associated with this action. We will now give an explicit formula
for $\mu$.

\begin{thm}
For any $m \in M$ and $(a,b) \in {\mathfrak g = su(2) \oplus su(2)}$
we have 
$$
\mu (a,b)\, =\, \frac{i}{4}\, \frac{f'(y)}{\sinh y}\, {\rm tr}(mm^{\dag}a -
m^{\dag}mb) 
$$
where $su(2)$ is identified with the space of $2\times 2$  
anti-hermitian matrices, and $y=y(m)$ is the function defined in
section $1$.
\end{thm}

\begin{proof}
Since $\omega = -\dd \alpha$, where $\alpha = \frac{i}{2} \D (f\circ
y)$ is a $G$-invariant 1-form on $M$, a well known result (th. 4.2.10
of \cite{A-M}) states that the moment map satisfies
\begin{gather}
\mu (m)\, [X]\, =\, \alpha_m (X^{\#})
\label{e6.1}
\end{gather}
for any $m\in M$ and $X \in {\mathfrak g}$, where $X^{\#}$ is the
vector field on $M$ generated by the one-parameter group of
biholomorphisms $\psi_{{\rm exp}(tX)} :M \rightarrow M$. Explicitly,
for any $(a,b) \in {\mathfrak g} = su(2) \oplus su(2)$ one can compute
\begin{gather}
(a,b)_{m}^{\#}\, =\, \frac{\dd}{\dd t} (e^{ta}m e^{-tb})_{t=0}\, =\,
am-mb \ ,
\label{e6.2}
\end{gather}
where we regard $T_{m}M \subset T_{m}GL(2,\C) \simeq M(2, \C)$.

On the other hand, for each $m \in M\setminus SU(2)$, the formula
$\tilde{y} := \cosh^{-1}(\frac{1}{2} {\rm tr} A^{\dag}A)$ gives a local
extension of $y$ to a neighbourhood in $M(2,\C)$ of $m$. Since $M$ is
a complex submanifold of $M(2,\C) \simeq \C^4$, it is then true that 
$\D (f\circ \tilde{y})|_{T_m M} = \D (f\circ y)_m$. Applying these
formulas we thus get
\begin{align}
\alpha_m [(a,b)^{\#}]\,& =\, \frac{i}{2}\, f'(y) \sum_{k=1}^{4} \frac{\D
\tilde{y}}{\D z_k}\ \dd z_k (am-mb)\, =  \nonumber \\ 
 & =\,  \frac{i}{2}\, f'(y)
\sum_{k=1}^{4} \frac{1}{2\sinh y}\, \bar{z}_k (m) \ \dd z_k (am-mb)\,
= \nonumber \\
 &  =\, \frac{i}{4}\, \frac{f'(y)}{\sinh y} \sum_{k,l=1}^{2} \bar{m}_{kl}\,
(am-mb)_{kl}\, =\, \frac{i}{4} \frac{f'(y)}{\sinh y}\, {\rm tr}(m^{\dag}am -
m^{\dag}mb)\ .
\end{align}
Since $\alpha$ and $(a,b)^{\#}$ are smooth on $M$, this formula can be
extended by continuity from $M\setminus SU(2)$ to $M$. It coincides with
the formula in the statement because of the cyclic property of the
trace.
\end{proof}

\begin{rem}
Although we will not reproduce the calculations here, a number of
properties of the moment map $\mu$ can be obtained quite
straightforwardly. For example, with respect to the norm on
$su(2)^{\ast }\oplus su(2)^{\ast}$ induced by the norm
$-{\rm tr}\,a^2$ on $su(2)$, one has
\begin{gather*}
\| \mu (m)\| ^2 \, =\, \frac{1}{4}\, f'(y(m))^2        \\
\mu (M)\, =\, \left\{ (a,b)\in su(2)^{\ast }\oplus su(2)^{\ast}
:\ \| a\| = \| b\|\ \in [0 , \frac{1}{2\sqrt{2}} f'(+\infty )[ \, \right\}
\end{gather*}
\end{rem}

The moment map obtained above associates to each $X \in {\mathfrak g}$
a function $\mu (\cdot)\, [X] \, \in C^{\infty}(M)$. In the framework
of geometric quantization this function is regarded as a classical
observable, and the quantization  procedure associates to it a certain
hermitian operator on the quantum Hilbert space. This correspondence is the
subject of the next section.

\section{Holomorphic quantization}

In this section we want to study the quantization of the classical
phase space $(M,\omega)$. We will use holomorphic quantization, which
is the simplest and most natural quantization procedure on a K\"ahler
manifold. Refinements such as the metaplectic correction will be left
out. For background material consult for example \cite{Wo}.
\footnote{ A warning about conventions: if $(M, \omega)$ is a
symplectic manifold and $H \in C^{\infty} (M)$, the definition of the
symplectic gradient vector field $X_H$ used in \cite{Wo} differs by a
sign from ours.}

We start with prequantization. Since the K\"ahler form $\omega =
\frac{i}{2} \D \Db  (f\circ y)$ is exact on $M$, the trivial
line-bundle $B := M\times \C$ with the canonical hermitian metric
$\left( (m, w_1) , (m, w_2) \right) = w_1 \bar{w_2}$ is a prequantum
bundle. Now consider the natural unitary trivialization of this bundle
$m \mapsto (m,1)$, and the connection $\nabla$ on $B$ defined by the
1-form
$$
\theta \, =\, \frac{1}{4 \hbar}\, (\Db - \D)(f\circ y)
$$
with respect to this trivialization. The curvature form of $\nabla$ is
$\dd \theta = -i\hbar^{-1} \omega $ and, since $\theta$ is pure
imaginary, the connection is compatible with the hermitian metric
$( \cdot , \cdot )$. Thus according to the definitions in \cite{Wo} $\left(
B , (\cdot ,\cdot) , \nabla \right)$ is the prequantum data.

The step from prequantization to quantization is made by choosing a
polarization on $M$. Since $M$ is K\"ahler the natural choice here is
the holomorphic polarization, that is, the polarization spanned by the
tangent vectors $\frac{\D}{\D z^k}$. With this choice, a section $m
\mapsto \varphi (m) = (m, \tilde{\varphi}(m))$ of $B$ is polarized iff
$\nabla^{0,1}\varphi =0 $, where $\nabla^{0,1}$ denotes the
anti-holomorphic part of the connection. But
\begin{gather}
\nabla^{0,1}\varphi\, =\, \Db \tilde{\varphi} + \tilde{\varphi}\,
\theta^{0,1}\, =\, \Db \tilde{\varphi} +
\frac{1}{4\hbar}\,\tilde{\varphi}\,\Db (f\circ y)\, =\, 0\  \iff 
\ \tilde{\varphi}\, =\, \phi\, e^{-(f\circ y) / 4\hbar }
\label{e7.1}
\end{gather}
where $\phi$ is any holomorphic function on $M$. Thus the space of
polarized sections of $B$ can be identified with the space of smooth
functions on $M$ of the form (\ref{e7.1}).

The final step to construct the quantum Hilbert space is to define an
inner product of polarized sections. This is done by the formula
\begin{gather}
\langle \varphi_1 , \varphi_2 \rangle\ =\ \int_M (\varphi_1 ,
\varphi_2)\ \epsilon\ =\ \int_M \phi_1 \bar{\phi_2} e^{ -(f\circ y)/
2\hbar} \ \epsilon  \ ,
\end{gather}    
where $\epsilon := (2 \pi \hbar)^{-3} \omega^3 /3!$
differs from the metric volume form on $(M,\omega)$ by the factor
$(2\pi \hbar)^{-3}$. The quantum Hilbert space of holomorphic
quantization, which we denote ${\mathcal H}_{HQ}$, is then defined as
the space of polarized sections of $B$ of finite $\langle \cdot ,
\cdot \rangle$-norm (see \cite{Wo}).

For a better understanding of this Hilbert space, one should get a
clearer picture of the holomorphic functions $\phi$ on $M$. This
picture is provided by the next proposition. Since its proof is rather
out of context and may easily be skipped, we defer the proof to the end of
the section.

\begin{thm}
Regard $M=SL(2,\C)$ as the zero set in $\C^4$ of the polynomial $D(z)=
z_1 z_4 -z_2 z_3 -1$. Then the natural restriction is an isomorphism
between the rings of holomorphic functions ${\mathcal O}(\C^4)/ J$ and
${\mathcal O}(M)$, where $J$ is the ideal of ${\mathcal O}(\C^4)$
generated by $D(z)$.
\end{thm}

In other words, this proposition states that every holomorphic
function on $M$ is the restriction of a holomorphic function on
$\C^4$, and furthermore two holomorphic functions on $\C^4$ restrict
to the same function on $M$ iff their difference is divisible by
$D(z)$. We thus get a characterization of holomorphic functions on $M$
in terms of entire functions on $\C^4$, which have a global power
series expansion and are generally much better understood.

$\ $

For the rest of this section we will look at operators on ${\mathcal
H}_{HQ}$. If $h \in C^{\infty}(M)$ is a classical observable,
geometric quantization associates to it an operator $\hat{h}$ on
${\mathcal H}_{HQ}$ defined by
\begin{gather}
\hat{h} (\varphi)\ =\ i\hbar\, \nabla_{Y_h}\, \varphi + h\,\varphi \qquad
\forall \varphi \in {\mathcal H}_{HQ} \ ,
\label{e7.2}
\end{gather} 
where $Y_h$ is the vector field on $M$ defined by $\iota_{Y_h} \omega
= \dd h$. This observable-operator correspondence does not always work,
however, because sometimes the resulting operator $\hat{h}$ takes
polarized sections into non-polarized ones. To prevent this, one
further demands that the flow of $Y_h$ should preserve the
polarization, i.e. the flow should be locally holomorphic. Thus
in principle not all observables $h \in C^{\infty}(M)$ can be
``quantized'' by this method. It can be shown, however, that if
this condition is fullfilled then $\hat{h}$ is a self-adjoint operator    
in the Hilbert space $({\mathcal H}_{HQ} , \langle \cdot , \cdot
\rangle)$ (see \cite{Wo} and review \cite{EMRV}).

We will now apply formula (\ref{e7.2}) to the observables coming from the
classical symmetries of $(M,\omega )$, that is to the functions $\mu^X
:= \mu (\cdot )\, [X] \, \in C^{\infty}(M)$ described in the previous
section. Notice that, by definition of moment map, for each $X \in 
{\mathfrak g}$ the vector field $Y_{\mu^X}$ is exactly $X^\#$ -- the
vector field generated by the one-parameter group of biholomorphisms
$\psi_{{\rm exp}(tX)}: M\rightarrow M$. In particular the flow of
$Y_{\mu^X}$, which is  $\psi_{{\rm exp}(tX)}$, preserves the
holomorphic polarization, and so formula (\ref{e7.2}) may be applied to $\mu^X$.

Putting together (\ref{e7.1}), (\ref{e7.2}) and (\ref{e6.1}) we get 
\begin{align}
\hat{\mu}^{X}\ \varphi\, & =\, i\hbar\, \nabla_{X^{\#}}
\varphi + \mu^X \varphi \, =\, i\hbar \left[ (\dd \varphi)(X^{\#}) +
\varphi \, \theta (X^{\#}) \right] + \alpha (X^{\#})\, \varphi\, =
\nonumber \\ 
& =\, i\hbar \left[ (\dd
\phi)(X^{\#}) - \frac{1}{4\hbar }\, \phi\, \dd (f\circ y) (X^{\# }) +
\frac{1}{4\hbar }\, \phi\, (\Db - \D )(f \circ y)\, (X^{\# }) \right]
e^{-(f\circ y)/ 4\hbar} + \nonumber \\
& \quad+ \frac{i}{2} \D (f\circ y)\,(X^{\#})\,
\varphi\   =\  i\hbar (\D \phi) (X^{\#})\, e^{-(f\circ y)/ 4\hbar } \ .
\label{e7.3}
\end{align}  

For an even more explicit formula, suppose $X = (a,b)\, \in
su(2) \oplus  su(2)$ and that $\phi \in {\mathcal O}(M)$
is the restriction of a certain  $\tilde{\phi} \in {\mathcal
O}(\C^4)$. Then using (\ref{e6.2}) and the fact that $M$ is a complex
submanifold of $\C^4$,
\begin{gather}
(\D \phi)_m (X^{\#}) = \sum_{k=1}^{4} \frac{\D \tilde{\phi}}{\D z^k}
(m)\, z^k (am-mb)\ ,
\label{e7.4}
\end{gather}
where $z^k (am-mb)$ stands for the entry $z^k$ of the matrix $am-mb$
under the identification $M(2,\C) \simeq \C^4$. Formulas (\ref{e7.3}) and (\ref{e7.4})
thus give an explicit description of the operator $\hat{\mu}^{X}$ on
${\mathcal H}_{HQ}$.

\begin{proof}[Proof of proposition 7.1]
This is a known consequence of textbook results.
 Let ${\mathcal A}_{\C^4}$ and  ${\mathcal A}_{M}$ be the sheaves of
  germs of holomorphic functions on $\C^4$ and $M$, respectively. By
  theorem 7.15 of \cite{Ho} these are coherent analytic
  sheaves. Furthermore, calling $\tilde{{\mathcal A}}_{M}$ the trivial
  extension to $\C^4$ of the sheaf ${\mathcal A}_{M}$ over $M$, it
  follows from theorems IV-D8 and VI-B5 of \cite{G-R} that
  $\tilde{{\mathcal A}}_{M}$ is still coherent analytic and has the
  same cohomology as ${\mathcal A}_{M}$.

Consider now the short sequence of sheaves over $\C^4$:
\begin{gather}
\begin{CD}
0 @>>> {\mathcal A}_{\C^4} @>\tilde{D}>> {\mathcal A}_{\C^4} @>r>> \tilde{{\mathcal
A}}_{M} @>>> 0 \ ,
\end{CD} 
\label{e7.5}
\end{gather}
where $\tilde{D}$ is the map induced by local multiplication by the
polynomial $D(z)$ and $r$ is defined by
\[
r|_U (f)\ =\ \left\{ 
\begin{aligned}
0 \ \ {\rm if} \ U \cap M = \emptyset \\
f|_{U\cap M} \ \ {\rm otherwise} 
\end{aligned} \right.
\qquad \qquad \in \ \Gamma(U, \tilde{{\mathcal A}}_{M})
\]
for every open set $U$ in $\C^4$ and every $f \in \Gamma (U, {\mathcal
A}_{\C^4})$. It is not difficult to check that (\ref{e7.5}) is in fact
an exact sequence. (Succinctly, $\tilde{D}$ is injective because the
stalks of ${\mathcal A}_{\C^4}$ are integral domains; $r$ is
surjective because $M$ is a complex submanifold of $\C^4$; ${\rm ker}\
r \subseteq {\rm im}\ \tilde{D}$ by the Nullstellensatz for germs of
varieties and the irreducibility of $D(z)$.)

We therefore obtain an exact sequence of cohomology groups 
\[
\begin{CD}
0 @>>> H^0 (\C^4 ,{\mathcal A}_{\C^4}) @>>> H^0(\C^4 ,{\mathcal
A}_{\C^4}) @>>> H^0(M, {\mathcal A}_{M}) @>>> 0 \ ,
\end{CD}
\] 
where we have used that $H^p(\C^4 , \tilde{{\mathcal A}}_{M}) \simeq
H^p(M , {\mathcal A}_{M})$ and that, by Cartan's theorem B
\cite[p. 243]{G-R}, $H^1 (\C^4 ,{\mathcal A}_{\C^4}) = 0$. Since the
zeroth cohomology groups 
are just the global sections of the respective sheaf and, under this
identification, the first and second maps are, respectively,
multiplication by $D(z)$ and the natural restriction, we finally
obtain that
\[
{\mathcal O}(M)\, =\, \Gamma (M,{\mathcal A}_{M} ) \, \simeq \,
\frac{\Gamma (\C^4 , {\mathcal A}_{\C^4})}{D(z)\cdot \Gamma (\C^4 ,
{\mathcal A}_{\C^4})} \, =\, {\mathcal O}(\C^4)/J \ .
\] 
\end{proof}

\section{Dimension of the Hilbert space}

In this last section of  the paper we will be concerned with the dimension
of the quantum Hilbert space associated with the K\"ahler manifold $(M,
\omega)$. More specifically, using the identification of the previous
section
\[
{ \mathcal H}_{HQ}\, \simeq \, \left\{ \phi \in {\mathcal O} (M) = {\mathcal
O}(\C ^4 ) / J   :  \int_M  |\phi |^2 e^{-(f \circ y)/ 2\hbar }\, \epsilon
<    +\infty  \right\} \ ,
\]
let ${\mathcal H}_{poly}$ be the subspace of  ${\mathcal H}_{HQ}$
consisting of the holomorphic functions that can be represented by
polynomials in $\C ^4$. Then we will be able to compute the dimension of
${\mathcal H}_{poly}$ in terms of  the K\"ahler potential
$f$. Furthermore, when
$(M, \omega )$ has finite volume $\Omega$ and ${\mathcal H}_{poly}$ has
finite dimension, we will show that ${\rm dim}_{\C} {\mathcal H}_{poly}
\sim  \Omega /(2\pi \hbar)^3$ as $\hbar \rightarrow 0^{+}$.  These results
are finally discussed in Questions 1, 2 and 3.

The main step towards proving the stated results is the following
proposition, which will be proved at the end of this section.

\begin{thm}  
Let $\phi$ be a holomorphic function on $M$ that can be represented by a
polynomial in $\C ^4$, of degree $l$, whose homogeneous term of  highest
degree is not divisible by $z_1 z_4 - z_2 z_3$. Then $\phi$ is in
${\mathcal H}_{poly}$ if and only if
\begin{gather}
\int_{0}^{+\infty} (\cosh y)^l \,  e^{-f(y) / 2\hbar}\, \frac{\dd}{\dd y}
[f'(y)]^3 \ \dd y  \ \ < \ +\infty \ .
\label{e8.1}
\end{gather}
\end{thm}

Using this proposition the dimension of ${\mathcal H}_{poly}$ can be   
computed quite straightforwardly. In fact, assuming  that $\{ l\in
{\mathbb N}: (\ref{e8.1})\ {\rm is}\ {\rm satisfied}  \}$ is not empty (which
will be shown to
be true when $(M,\omega)$ has finite volume), and calling $m\in  {\mathbb
N} \cup \{ +\infty \}$ the maximum of this set, we have:

\begin{cor}
The complex dimension of  ${\mathcal H}_{poly}$ is  $\frac{1}{6} (m+1)
(m+2) (2m+3)$.
\end{cor}

\begin{proof}  
Let $P_l  \subset {\mathcal O}(\C^4)$ be the subspace of homogeneous     
polynomials of degree $l$, and $P_{\le m}$ the space $\bigoplus_{0\le l  
\le m} P_l$. Calling $\chi : {\mathcal O}(\C^4) \rightarrow {\mathcal
O}(M)$ the natural homomorphism, let also $\chi_{|}$ be the restriction of
$\chi$  to $P_{\le m}$.

By proposition $8.1$ we have that ${\mathcal H}_{poly} =  \chi (P_{\le m})$,
thus
\[
{\rm dim}\,{\mathcal H}_{poly}\, =\, {\rm dim}(P_{\le m})  - {\rm dim}({\rm     
ker}\, \chi_| )\, =\, {\rm dim}(P_{\le m}) - {\rm dim}(P_{\le m} \cap {\rm   
ker}\, \chi ) \ .
\]
But proposition $7.1$ states that  ${\rm ker}\ \chi $ is the ideal in ${\mathcal
O}(\C^4)$ generated by $z_1 z_4 -z_2 z_3 -1$, and so it is clear that the
linear map
\[
P_{\le m-2}\, \rightarrow\, P_{\le m} \cap {\rm ker}\,\chi \qquad
,\qquad  Q(z) \mapsto (z_1 z_4 - z_2 z_3 -1) \cdot Q(z)
\]
is an isomorphism. Hence ${\rm dim}( P_{\le m} \cap {\rm ker}\, \chi ) = {\rm
dim}(P_{\le m-2})$, and $ {\rm dim}\,{\mathcal H}_{poly} = {\rm dim} (P_m
\oplus P_{m-1})$. But it's a well known combinatorial fact that
dim$P_m$ -- the number of ways of choosing 4 non-negative integers
whose sum is $m$ -- is $\binom{3+m}{3}$, and the result follows directly. 
\end{proof}

In practice, by looking at the asymptotics of the potential $f$, it is
usually not difficult to compute the integer $m$.

\begin{exa}
The lump metric on $M$ studied in section $5$ has a K\"ahler form
$\pi^{\ast}  \omega = \frac{i}{2} \D \Db (f \circ y)$, with $f(y) =
\pi y \coth
y$, and finite volume $\Omega = \pi^6 /3$. From the asymptotics
\[
f(y) = \pi y\, [ 1+ 2 e^{-2y} + 2e^{-4y} + O (e^{-6y})]  \ \ \ \ \ \ \ {\rm
as} \ \ \ y \rightarrow +\infty
\]
one gets that
\[
(\cosh y)^l \, e^{-f(y) / 2\hbar}\, \frac{\dd}{\dd y} [ f'(y)]^3 \,
=\,  O (y\, 
e^{(-\pi /2\hbar  + l -2)y})   \ \ \ \ \ \ {\rm as} \ \ \ \ y \rightarrow
+\infty  \ ,
\]
and so $m = \max \{ l \in {\mathbb N}: l < 2+ \pi / 2 \hbar    \}$.
\end{exa}

An interesting feature of this example is that, if we let $\hbar
\rightarrow  0^+ $, then  $m \sim \pi / 2\hbar$, and by corollary $8.1$ we  
obtain
\[
{\rm dim}_{\C}\,{\mathcal H}_{poly}\ \sim\ \frac{m^3}{3}\ \sim\  \frac{\pi ^3}{24
\hbar ^3} = \frac{\Omega}{(2\pi \hbar)^3} \ .
\]
This is exactly the answer expected in semi-classical quantum mechanics for
the quantization of  a phase space of volume $\Omega$ and real dimension 6
\cite{Ku}. Before discussing the significance of this coincidence, we will
first show that this property is more general, and is in fact valid for   
all the $G$-invariant K\"ahler  metrics on $M$ of finite volume.

\begin{thm}
Suppose $ \omega = \frac{i}{2} \D \Db (f \circ y)$ is the
K\"ahler form of a metric on $M$ of finite volume $\Omega$. Then the
constant $m = m(f, \hbar )$ satisfies
\[
\frac{(3\Omega)^{1/3}}{2\pi \hbar } + k -1\ \le\ m\ \leq
\ \frac{(3\Omega)^{1/3}}{2\pi \hbar } + k  \ ,
\]
where
\[
k\, =\, k(f)\, :=\,
 \sup \left\{  \lambda \in \R_{0 }^{+}: \int_{0}^{+\infty }
e^{\lambda y}\, \frac{\dd}{\dd y} [f'(y)]^3 \, \dd y\ \ {\rm is}\ {\rm
finite}\right\}
\ \in \ [0, +\infty ]  \ .
\]
\end{thm}

\begin{proof}
By corollary $4.2$ we have that
\begin{gather}
\Omega\, =\, \frac {\pi ^3}{3} \int_{0}^{+\infty} \frac{\dd}{\dd y} [f'(y)]^3
\, \dd y\,  =\, \frac{\pi ^3}{3} \lim_{y \rightarrow +\infty} [ f'(y)]^3 \ ,
\label{e8.2}
\end{gather}
thus the finite volume condition implies that $k \ge 0$ and that $D :=
\lim_{ y \rightarrow +\infty}  f'(y)$ is a  positive finite number. By
L'H\^opital's rule we also get that  $\lim_{ y \rightarrow +\infty}
y^{-1}f(y)  = D$. It then follows that, for $\beta$ real,
\begin{gather}
\lim_{ y \rightarrow +\infty }\  (e^{-y} \cosh y)^l \, \exp \left\{ \left[\beta -
\frac{1}{2\hbar} \left(\frac{f(y)}{y} - D\right)\right]\, y \right\} \ =\  \left\{
\begin{aligned}
+\infty\ {\rm if} \ \beta  > 0   \\   0\ \ {\rm if}\ \beta  <  0
\end{aligned} \right. \ .
\label{e8.3}
\end{gather}
Now, by definition, $m$ is the biggest of the integers $l\in {\mathbb N}$
such that
\begin{gather}
\int_{0}^{+\infty} (\cosh y)^l\,  e^{-f(y) / 2\hbar}\, \frac{\dd}{\dd y}
[f'(y)]^3 \ \dd y
\label{e8.4}
\end{gather}
converges. This integral, however, is the same as
\[
\int_{0}^{+\infty} \left\{ (e^{-y} \cosh y)^l  \exp \left[ \beta -
\frac{1}{2\hbar}\left(\frac{f(y)}{y} - D\right)  y \right] \right\}\, e^{(- \beta + l -
D/2\hbar )\, y }\, \frac{\dd}{\dd y} [f'(y)]^3 \ \dd y \ .
\]
If  $l - D/2\hbar > k $, then choosing  $\beta$ in the interval $]0\, ,\, l-
D/2\hbar - k [$ and using  (\ref{e8.3}) and the definition of $k$, it is clear
that (\ref{e8.4}) diverges; thus $m \le k + D/ 2\hbar $.  If  $l - D/2\hbar < k $,
then choosing  $\beta$ in $] l- D/2\hbar - k\, ,\, 0[$, the same arguments    
show that (\ref{e8.4}) converges; thus  $m \ge k + D/ 2\hbar -1$.  Since from (\ref{e8.2})
we have $D = (3\Omega)^{1/3} / \pi$, the proposition is proved.
\end{proof}

\begin{cor}
Given any $G$-invariant K\"ahler metric on $M$ of finite volume $\Omega$,
let $\omega = \frac{i}{2} \D \Db (f \circ y)$ be its K\"ahler form. Then
the associated quantum space ${\mathcal H}_{poly}$ is finite-dimensional
iff $k(f)$ is finite. In this case ${\rm dim}_{\C}{\mathcal H}_{poly} \sim
\Omega /(2\pi \hbar)^3$ as $\hbar \rightarrow 0^+ $.
\end{cor}

\begin{proof}
It follows directly from corollary $8.2$ and proposition $8.4$.
\end{proof}

\begin{rem}
The lump metric is an example with finite volume and finite-dimensional
${\mathcal H}_{poly}$. There are however metrics of finite volume with
infinite dimensional  ${\mathcal H}_{poly}$. For example, define $f'(t)$ as
any extension of $t \mapsto (1 - e^{-t^2}) \ \ ,\ t \in [1, +\infty[ $, to
a smooth odd function on $\R$ with everywhere positive first derivative.
Then calling $f$ any primitive of $f'$, the metric on $M$ with K\"ahler
potential $f \circ y$ has the desired  property.
\end{rem}

\subsection{Discussion}

The asymptotic value $\Omega /(2\pi \hbar)^3$ for the complex dimension of
the quantum Hilbert space is both physically and geometrically quite
significant. 
Physically because semi-classical statistical mechanics predicts that
if you quantize a classical system with $n$ degrees of freedom, thus
with $2n$-dimensional phase space, you should get one independent
quantum state of the system for each cell of volume $(2\pi \hbar)^n$
on the phase space \cite{Ku}. Hence if the phase space has finite
volume $\Omega$, one gets $\Omega /(2\pi \hbar)^n$ independent quantum states.

The geometrical significance, on the other hand, arises from the fact
that this asymptotic value is expected when the base manifold is
compact, but not ``a priori'' for the non-compact $M=SL(2,\C)$. The
compact result is a direct consequence of the Hirzebruch-Riemann-Roch
formula and the Kodaira vanishing theorem \cite{Hir}, and can be
stated as follows.
 
\vspace{4mm}
{\it On a compact K\"ahler $2n$-manifold of volume $\Omega$, the Hilbert
space of holomorphic quantization has finite complex dimension, and this grows
asymptotically as $\Omega / (2\pi \hbar)^n$ when $\hbar \rightarrow
0^+$.}
\vspace{3mm}

 We are thus led to the following questions.

\begin{que1}
Is there a version of the above result for the non-compact case ? 
\end{que1}
Our results suggest that there is, since $SL(2,\C)$ is not compact and
some of the $G$-invariant metrics for which corollary $8.5$ holds -- for
example the lump metric -- cannot be compactified.

Another example is the manifold $\C$ with any $U(1)$-invariant
K\"ahler metric $g$. In this case the K\"ahler form also has a
$U(1)$-invariant global potential -- which we call $\rho$ -- and
${\mathcal H}_{poly}$ is the space of square-integrable complex
polynomials on $\C$ with respect to the volume form $\exp{(- \rho
/2\hbar)}\, \frac{i}{2} \D \Db \rho$. A simplified version of the method
used in this paper then shows that, if the volume $\Omega$ of $(\C
,g)$ and the dimension of ${\mathcal H}_{poly}$ are both finite, we
also have 
\[
{\rm dim}_{\C}{\mathcal H}_{poly}\ \sim \ \Omega /(2\pi \hbar) \qquad
{\rm as}\ \ \hbar \rightarrow 0^+ \ .
\]  
In an optimistic spirit, we are thus led to formulate the following
question. 
\begin{que2}
Let $S$ be a closed complex submanifold of $\C^N$ (i.e. a Stein
manifold) of complex dimension $n \leq N$, and let $\omega$ be any
K\"ahler form on $S$. Since $S$ is Stein, $\omega$ has a global
potential $\rho \in C^{\infty}(S;\R)$, and we can define
\[
{\mathcal H}_{poly}\, :=\, \left\{ \phi \in \C [z_1 , \ldots , z_N] :
\int_S  |\phi|^2  e^{- \rho /2\hbar} \omega^n \ < \ +\infty
\right\}\ .
\]
Then, if the volume $\Omega$ of $(M, \omega)$ and the dimension of
${\mathcal H}_{poly}$ are both finite, is it always true that ${\rm
dim}_{\C}{\mathcal H}_{poly} \, \sim \, \Omega / (2\pi \hbar)^n$ as
$\hbar \rightarrow 0^+ $ ? 
\end{que2}
Having in mind our examples, it is also possible that the result only
holds for algebraic submanifolds of $\C^N$. 
Another point which would be worth to clarify is the relation between
the spaces ${\mathcal H}_{poly}$ and ${\mathcal H}_{HQ}$. 

\begin{que3}
In our example of $SL(2,\C )$ with a $G$-invariant K\"ahler metric of
finite volume, is it true that whenever ${\mathcal H}_{poly}$ is
finite-dimensional we have ${\mathcal H}_{poly} = {\mathcal H}_{HQ}$ ?
And for more general Stein manifolds of finite volume ?
\end{que3}
Recall that the finite-dimensionality of ${\mathcal H}_{poly}$ comes
from the fact that the only polynomials on $\C^4$ which are integrable
on $SL(2, \C)$, are the ones of degree smaller than a certain
constant. The above question asks if, in this case, the entire
non-polynomial functions on $\C^4$ are automatically non-integrable on
$SL(2, \C)$.
It is plausible that the answer is yes, since entire non-polynomial
functions have very high growth rates in certain directions. However,
if the answer is no, then perhaps in this case it is wiser to take
${\mathcal H}_{poly}$ as the quantum Hilbert space, instead of the
traditional ${\mathcal H}_{HQ}$. The finite-dimensionality of
${\mathcal H}_{poly}$ ensures completeness and corollary $8.5$ supports
this choice.

\subsection{Proof of proposition $8.1$}
  
\begin{proof}[The ``only if'' statement] 
We have to show that
\begin{gather}
\int_M |\phi |^2  e^{-(f\circ y)/ 2\hbar} \, \epsilon \ \ <\  +\infty
\label{e8.5}
\end{gather}
implies condition (\ref{e8.1}).  Notice first that if (\ref{e8.5}) is
satisfied, the 
``change of variables'' theorem guarantees that for any of the $G$-action  
biholomorphisms $\psi_g  : M \rightarrow M$,
\[
\int_M |\phi |^2  e^{-(f\circ y)/ 2\hbar} \, \epsilon \ = \ \int_M |\phi
\circ \psi_g |^2  e^{-(f\circ y)/ 2\hbar} \, \epsilon \ ,
\]
where we have used the $G$-invariance of the volume form $\epsilon$. So
using the invariant (Haar) integral on $G$ to average over the group, we
get that
\begin{gather}
\int_M |\phi |^2  e^{-(f\circ y)/ 2\hbar } \, \epsilon \  = \int_M \left(  
\int_{g\in G} |\phi \circ \psi_g |^2 \right)  e^{-(f\circ y)/ 2\hbar } \,  
\epsilon \ .
\label{e8.7} 
\end{gather}
Now regard $\phi (z_1 , \ldots ,z_4 )$ as a polynomial  on $\C^4$, and    
write $\phi = \phi_0 + \cdots + \phi_l $, where $\phi_k$ is homogeneous of
degree $k$. As in Appendix A, consider also the natural extension of
the $G$-action $\psi$ to the manifold $\C^4 \subset M$. We then have
\[
\int_{g\in G} |\phi \circ \psi_g  (z) |^2 \ = \ \sum_{k,j =0}^{l}
\int_{g\in G} (\bar{\phi_k}\phi_j ) \circ \psi_g (z)\ \ ,
\]
where each term of the sum is a smooth $G$-invariant function on $\C^4$.
In particular, using the notation and proposition A$.4$ of Appendix A,
each of these terms may be written as $F_{kj}(x(z), w(z))$, where $F_{kj}
: {\mathcal B} \rightarrow \C$ is continuous, $x(z) = (|z_1|^2 + \cdots +
|z_4|^2 )/2 $ and $w(z) = z_1 z_4 - z_2 z_3 $. On the other hand, going
back to the definition of $\psi$, we see that each component of $\psi_g
(z)$ is just a linear combination of  the components of $z$, and so in
fact we must have
\[
F_{kj}(x(z), w(z))\, =\, \int_{g\in G} (\bar{\phi}_k\phi_j ) \circ \psi_g (z)
\, =\, \sum c_{i_1 \cdots i_k n_1 \cdots n_j}\, \bar{z}_{i_1} \cdots
\bar{z}_{i_k} z_{n_1} \cdots z_{n_j}\ \  .
\]
From this formula it is clear that, for any $\lambda \in \R^{+}_{0}$,   
\[
F_{kj}(\lambda^2 x(z),\, \lambda^2 w(z))\, =\, F_{kj}(x(\lambda z),\, w(\lambda
z))\,  =\, \lambda^{k+j} F_{kj}(x(z),\, w(z)) \ ,
\]
and in particular
\[
x^{-l}\, F_{kj}(x,1)\, =\, x^{-l + (k+j)/2}\, F_{kj} (1,\, x^{-1})  \ .
\]
Since $k,j \le l$, using the continuity of  $F_{kj}$ we then obtain that  
\[
\lim_{x\rightarrow +\infty} x^{-l}\, F_{kj}(x,\,1)\, =\,  \lim_{x\rightarrow     
+\infty} x^{-l + (k+j)/2}\, F_{kj} (1,\, x^{-1})\, =\,  \delta_{lk}\, \delta_{lj}\,
F_{ll}(1,0) \ .
\]
Defining
\[
h(x(z) ,\, w(z) )\, :=\, \int_{g\in G} |\phi \circ \psi_g  (z) |^2 \, =\, \sum_{k,j
=0}^{l} F_{kj}(x(z),\, w(z)) \ ,
\]
we therefore have that
\begin{gather}
\lim_{x\rightarrow +\infty} x^{-l}\, h(x,1)\, =\, \sum_{k,j =0}^{l} \lim_{x  
\rightarrow +\infty} x^{-l}\, F_{kj}(x , 1)\, =\, F_{ll} (1,0) \ \ .
\label{e8.6}
\end{gather}

Now, it will be shown in lemma $8.6$ that $0 < F_{ll}(1,0) < +\infty $, and
so (\ref{e8.6}) implies that there is a constant $c> 0 $ such that $h(x,1) > c\,
x^l$ for $x$ big enough. On the other hand, using (\ref{e8.7}), proposition $4.1$
of section $4$, and that $w(z) =1$ and $x(z) = \cosh (y(z))$  for $z\in
M$, we have
\begin{align*}
\int_M |\phi |^2 \, e^{-(f\circ y)/ 2\hbar} \ \epsilon \ \,& = \ \int_{z\in M}
h(x(z),\, w(z))\,  e^{-(f\circ y)/ 2\hbar} \ \epsilon\ = \\
& =\, \frac{\pi^3}{3(2\pi
\hbar)^3} \int_{0}^{+\infty} h(\cosh y , 1)\,  e^{-f(y) / 2\hbar}\,
\frac{\dd}{\dd y} [f'(y)]^3 \ \dd y \ \ge \\
& \ge \  {\rm const.} +  \frac{c}{24 \hbar^3 } \int_{0}^{+\infty} (\cosh
y )^l\,  e^{-f(y) / 2\hbar}\, \frac{\dd}{\dd y} [f'(y)]^3 \ \dd y \ ,
\end{align*}
where const. is some finite real number. From this inequality it is clear
that (\ref{e8.5}) implies condition (\ref{e8.1}) of proposition $8.1$.
\begin{lem}
The constant $F_{ll}(1,0)$ is in $]0 , +\infty[$.
\end{lem}
\begin{proof}
As we have seen above $F_{ll}: {\mathcal B} \rightarrow \C $ is continuous
and, by definition,
\begin{gather}
F_{ll}(x(z), w(z))\, =\,  \int_{g\in G} |\phi |^2 \circ \psi_g  (z) \ .
\label{e8.8}
\end{gather}  
Since $(1,0) \in {\mathcal B}$ (see Appendix A), $F_{ll}(1,0)$ is a   
well-defined finite number, and from (\ref{e8.8}) it is clearly non-negative. Now
call ${\mathcal V} := \{ z\in \C^4 : z_1 z_4 - z_2 z_3 =0  \}$, and let
$q$ be any point in ${{\mathcal V} \setminus \{ 0 \}}$. Since $q' :=
x(q)^{-1/2} q \, \in {\mathcal V}$ and $x(q') = 1$, we have
\[
\int_{g\in G} |\phi_l  \circ \psi_g  (q') |^2 \ = \ F_{ll} (x(q'),
w(q'))\, =\, F_{ll}(1,0)\ \ .
\]
Hence, if $F_{ll} (1,0) = 0$, we get that $\phi_l \circ \psi_g (q') = 0 $
for any $g \in G$, and in particular $\phi_l (q') = 0 $. But $\phi_l$ is
homogeneous, and so also $\phi_l (q) = 0$. From the arbitrariness of $q$
it follows that $\phi_l$ vanishes on ${\mathcal V}$ -- the zero set
of the irreducible polynomial $z_1 z_4  - z_2 z_3 $. Finally from
Hilbert's Nullstellensatz we conclude that $\phi_l $ is divisible by $z_1 
z_4 - z_2 z_3$. This contradicts the hypothesis of proposition $8.1$, and
therefore $F_{ll}(1,0) > 0$.
\end{proof}
  
\begin{proof}[The ``if'' statement]
As before, write $ \phi = \phi_0  + \cdots + \phi_l $. Then $ |\phi |^2 
\le |\phi_0|^2   + \cdots + |\phi_l |^2 \ . $ Since $x(z) = (|z_1|^2  +
\cdots + |z_4|^2 )/2$, we have that $|z_k| \le \sqrt{2 x(z)}$ for all $z$.
But $\phi_j $ is homogeneous of degree $j \le l$, thus
\[
| \phi_j (z)|^2\ \le\  c_j\, (2x(z))^j\ \le\ c_j\, (2x(z))^l \ \ \ ,\ \ \ z \in
\C^4 \ ,
\]
for some positive constants $c_j$. Calling $c = \sum_{j=0}^l c_j$, 
using proposition $4.1$
of section $4$ and that $x(z) = \cosh y(z)$ for $z \in M$, we finally get
\begin{gather*}
\begin{split}
\int_M |\phi |^2  e^{-(f\circ y)/ 2\hbar} \, \epsilon \ \le \
\sum_{j=0}^{l} \int_M |\phi_j |^2  e^{-(f\circ y)/ 2\hbar} \, \epsilon\
\le \  \int_{z\in M} c\, (2x(z))^l  e^{-(f\circ y)/ 2\hbar} \, \epsilon \
= \\ =\ \frac{c\, 2^l \,\pi^3}{3(2\pi \hbar )^3} \int^{+\infty}_{0} (\cosh y
)^l\,  e^{-f(y) / 2\hbar}\, \frac{\dd}{\dd y} [f'(y)]^3 \ \dd y \ .
\end{split}
\end{gather*}
From this inequality it is clear that condition (\ref{e8.1}) of proposition $8.1$
implies (\ref{e8.5}), and so $\phi \in  {\mathcal H}_{poly}$.
\end{proof}
\renewcommand{\qedsymbol}{}
\end{proof}

\vskip 25pt
\noindent
{\bf Acknowledgements.}
I would like to thank Prof. N. S. Manton for many helpful discussions
and Prof. B. Totaro for some comments regarding proposition 7.1.
I am supported by \lq{\sl Funda\c{c}\~ao para a Ci\^encia e
Tecnologia}\rq, Portugal, through the research grant
SFRH/BD/4828/2001.

\newpage
\noindent
{\Large {\bf Appendix A}}\\
$\ $\\
$\ $

In this appendix we study the action $\psi$ of the group $G := SU(n)
\times SU(n)$ on the manifold $M(n, \C) \simeq \C ^{n^2}$ of complex
$n\times n$ matrices defined by
\begin{gather}
\psi : G\times M(n,\C ) \rightarrow M(n, \C)\quad ,\quad (U_1 ,U_2 ,A)
\mapsto U_1 A U_2^{-1}\ .
\end{gather} 
The results obtained are used in sections $2$ and $8$.

According to \cite[p. 396]{Kn} every matrix $M \in GL(n, \C)$ can be
decomposed in the form $M = KAK'$, where $K, K' \in U(n)$ and $A$ is
real diagonal with positive entries in the diagonal. Notice that
multiplying $K$ and $K'$ by permutation matrices, if necessary, we
may assume that the diagonal entries of $A$ do not decrease with the
row index.

\begin{lema1}
Every matrix $M \in M(n, \C)$ may be decomposed in the form $ M = U_1
A U_2 e^{i \theta}$, where $U_1 ,U_2 \in SU(n), \, \theta \in \R$, and
$A$ is a real diagonal matrix with non-negative diagonal entries which do not
decrease with the row index.
\end{lema1}

\begin{proof}
Given $M \in M(n, \C)$ there is a sequence $\{M_j\}$ in $GL(n, \C)$ with
$M_j \rightarrow M$. Using the decomposition described above, for each
$M_j$ we have 
$$
M_j \, =\, K_j A_j K'_j \ .
$$
Since the sequences $\{K_j \}$ and $\{K'_j \}$ are in the compact
group $U(n)$, there are convergent subsequences $K_{j_l} \rightarrow
K$ and $K'_{j_l} \rightarrow K'$ when $l\rightarrow +\infty$, where
$K, K' \in U(n)$. Defining 
\[
A\, :=\, K^{\dag} M (K')^{\dag}\, =\, \lim_{l \rightarrow +\infty}
(K_{j_l})^{\dag} M_{j_l}(K'_{j_l})^{\dag}\, =\, \lim_{l \rightarrow +\infty} A_{j_l}\ ,
\]
the fact that $A_{j_l}$ is diagonal with positive ordered diagonal
entries, implies that $A$ is diagonal with non-negative ordered
diagonal 
entries; furthermore $KAK' = M$. Since $K,K' \in U(n)$, they can
always be written as matrices in $SU(n)$ times a phase, and this ends
the proof.
\end{proof}

We will now find functions on $M(n, \C)$ which separate the orbits of
$\psi$, and hence can be used as coordinates in the space of
orbits. For this define the polynomials $P_j$ on $M(n, \C)$ by
$$
\det (B + \lambda I)\, =\, \sum_{j=0}^{n} \lambda^j \, P_j (B)\ .
$$
We then have:

\begin{propa2}
Two matrices $M,N \in M(n,\C)$ lie in the same orbit of $\psi$ if and
only if $P_j (M^{\dag}M) = P_j (N^{\dag}N)$ for $1\leq j\leq n$ and
$\det N = \det M$.
\end{propa2}

\begin{proof}
If $N$ and $M$ are in the same orbit, i.e. $N=U_1 M U_2$ for some $U_1
,U_2 \in SU(n)$, then $N^{\dag}N= U_2^{\dag} M^{\dag} M U_2$, and the
stated conditions are clearly satisfied.

Conversely, suppose that $P_j (N^{\dag}N) = P_j (M^{\dag}M)$ for
$1\leq j\leq n$ and $\det M = \det N$. Then $N^{\dag}N$ and
$M^{\dag}M$ have the same characteristic polynomial, and hence the
same eigenvalues. On the other hand, from lemma A.1 we have the
decompositions 
\begin{gather}
M\, =\, U'\, {\rm diag}(\lambda_1 ,\ldots ,\lambda_n)\, U
e^{i\theta}\quad {\rm and}\quad N\, =\, \tilde{U}'\, {\rm diag}
(\tilde{\lambda}_1 , \ldots , \tilde{\lambda}_n ) \, \tilde{U}\, e^{i\beta}\ ,
\label{eA1}
\end{gather}
so
$$
M^{\dag}M\, =\, U^{\dag}\, {\rm diag}(\lambda_1^2 , \ldots , \lambda_n^2)\, U
\ \ \ {\rm and}\quad N^{\dag}N\, =\, \tilde{U}^{\dag}\, {\rm
diag}(\tilde{\lambda}_1^2 , \ldots , \tilde{\lambda}_n^2)\, \tilde{U} 
$$
have eigenvalues $\{ \lambda_1^2 ,\ldots ,\lambda_n^2\}$ and
$\{\tilde{\lambda}_1^2 , \ldots , \tilde{\lambda}_n^2 \}$,
respectively. Since lemma A.1 also guarantees that the $\lambda_j ,
\tilde{\lambda_j}$ are non-negative and ordered, we conclude that
$\lambda_j = \tilde{\lambda_j}$ for $1\leq j \leq n$. Hence
\[
N\, =\, \tilde{U'}(U')^{-1}M U^{-1} \tilde{U}\, e^{i(\beta -\theta)}\ .
\]
If $\det M = \det N \ne 0$, then taking the determinant of the above
equation we get that $\det (e^{i(\beta -\theta)} I) = 1$, and so
$e^{i(\beta -\theta)} I$ is in $SU(n)$. This shows that $N=U_1 MU_2$
for some $U_1 ,U_2 \in SU(n)$.

If $\det M = \det N = 0$, then (\ref{eA1}) implies that the product of the
$\lambda_j$ is zero, therefore $\lambda_1 = \tilde{\lambda}_1 =0$,
because the $\lambda_j$ are non-negative and ordered. Defining
$$
\Lambda\, :=\, {\rm diag}(e^{i(\theta -\beta)(n-1)}, e^{i(\beta -\theta)},
\ldots, e^{i(\beta -\theta)} )\ \ \in\ SU(n)
$$ 
we then get
$$
N\,=\, \tilde{U'}\,{\rm diag}(\lambda_1, \ldots ,\lambda_n)\,\tilde{U}\,
e^{i\beta}\, =\, \tilde{U'}\,{\rm diag}(\lambda_1, \ldots ,\lambda_n)\, \Lambda
\tilde{U}\, e^{i\theta}\, =\,   \tilde{U'}(U')^{-1}M U^{-1} \Lambda
\tilde{U} \ , 
$$
which shows that, also in this case, $N=U_1 MU_2$ for some $U_1 ,U_2
\in SU(n)$. 
\end{proof}

These results are now going to be used in the study of $G$-invariant
functions on $M(2,\C)$ and $SL(2,\C)$. Define the smooth map
\[
\beta : M(2,\C) \simeq \C^4 \rightarrow \R \times \C \quad ,\quad
\beta(z)= (x(z),w(z)) \ ,
\]
where
\[
x(z)\, =\, \frac{1}{2} (|z_1|^2 +\cdots +|z_4|^2)\quad {\rm and}\quad
w(z)\, =\, z_1 z_4 -z_2 z_3 \ .
\]
It follows from the proposition above that two points in $M(2,\C)$ lie
in the same $\psi$-orbit iff they have the same image by $\beta$. In
particular any $G$-invariant function $\tilde{h}$ on $M(2,\C)$ may be
written $\tilde{h} = h \circ \beta$, where $h$ is some function
defined on the image of $\beta$. We will now show that the continuity
of $\tilde{h}$ implies the continuity of $h$ -- a result used in
section $8$.

\begin{lema3}
The image of $\beta$ is ${\mathcal B} := \{ (a,u) \in \R \times \C : a
\ge |u| \}$.
\end{lema3}

\begin{proof}
From the identity 
$$
x(z)^2 - |w(z)|^2 \, =\, \frac{1}{4} (|z_1|^2+|z_2|^2-|z_3|^2-|z_4|^2)^2 +
|z_1 \bar{z_3} + z_2 \bar{z_4}|^2 \, \ge \, 0
$$
it follows that $x(z) \ge |w(z)|$, thus the image of $\beta$ is
contained in ${\mathcal B}$.

Conversely, defining $g:{\mathcal B} \rightarrow \C^4$ by
\begin{gather}
g(a,u)\, =\, \left( u(a + \sqrt{a^2 - |u|^2})^{-1/2}, 0, 0, (a + \sqrt{a^2
- |u|^2})^{1/2} \right)\ ,
\label{eA.2}
\end{gather}
one can easily check that $\beta \circ g (a,u) = (a,u)$, and so
${\mathcal B}$ contains the image of $\beta$.
\end{proof}

\begin{propa4}
Let $X$ be a topological space, ${\mathcal V}$ a subset of the image
of $\beta$, and $h: {\mathcal V} \rightarrow X$ a map such that $h
\circ \beta$ is continuous on $\beta^{-1} ({\mathcal V})$. Then $h$ is
continuous. 
\end{propa4}

\begin{proof}
Consider the map $g:{\mathcal B} \rightarrow \C^4$ defined in
(\ref{eA.2}). This map is clearly continuous on ${\mathcal B} \setminus
\{(0,0)\}$ and, for $(a,u)$ approaching the origin from this set,
$$
\lim_{(a,u)\rightarrow (0,0) }
\left| \frac{u}{\sqrt{a + \sqrt{a^2 - |u|^2}}} \right| \,\leq \,
\lim_{(a,u)\rightarrow (0,0) }
\frac{a}{\sqrt{a}}\, =\, 0 \ .
$$
Thus $g$ is also continuous at $(0,0)$ and vanishes at this
point. Finally, since
$$
h(a,u)\, =\, (h \circ \beta) \circ g (a,u) \quad {\rm for}\ {\rm all}\ \ (a,u)\in
{\mathcal V}\ , 
$$ 
we conclude that the continuity of $h\circ \beta$ implies the
continuity of $h$.
\end{proof}

Now suppose we restrict the action $\psi$ of $G$ to the submanifold
$SL(2,\C) \subset M(2,\C)$. Since the function $w(z)$ is identically
$1$ on $SL(2,\C)$, we have that two points in this submanifold lie in
the same $\psi$-orbit iff they have the same image by $x(z)$. From
lemma A.3 it follows that $x(SL(2,\C)) = [1,+\infty[$, and since
$\cosh^{-1}$ is injective on this interval, we have that $y:=
\cosh^{-1}\circ\, x$ also separates orbits in $SL(2,\C)$. We can now
prove proposition $2.1$ of section $2$.

\begin{proof}[Proof of proposition $2.1$] 
From the paragraph above, it is clear that any smooth $G$-invariant
function $\tilde{f}$ on $SL(2,\C)$ may be written as $\tilde{f} =
f\circ y$, for some unique $f :[0,+\infty[ \rightarrow \R$. Now
consider the smooth map $h: \R \rightarrow SL(2,\C)$ defined by
\[
h(t)\, =\, \begin{bmatrix}
       \cosh (t/2)  &  \sinh (t/2) \\
       \sinh (t/2)  &  \cosh (t/2)
       \end{bmatrix}\ .
\]
One can easily check that $y\circ h(t) = t$ for $t \ge 0$, hence
$$
f(t)\, =\, (f\circ y)\circ h(t)\, =\, \tilde{f}\circ h(t)\ \ ,\ \ t\ge 0\ ,
$$
which implies that $f$ is smooth. Note also that $h$ is defined on $\R$,
and from the $G$-invariance of $\tilde{f}$ we get $\tilde{f} \circ h
(-t) = \tilde{f} \circ h (t)$; thus $f$ can be extended to an even
function on $\R$.
\end{proof}

\end{document}